\documentclass{jfm}

\usepackage{graphicx}
\usepackage{natbib}
\usepackage{epstopdf}
\usepackage{float}
\usepackage{amssymb}
\usepackage{amsbsy}
\usepackage{color}
\usepackage{amsmath}

\ifCUPmtlplainloaded \else
  \checkfont{eurm10}
  \iffontfound
    \IfFileExists{upmath.sty}
      {\typeout{^^JFound AMS Euler Roman fonts on the system,
                   using the 'upmath' package.^^J}%
       \usepackage{upmath}}
      {\typeout{^^JFound AMS Euler Roman fonts on the system, but you
                   dont seem to have the}%
       \typeout{'upmath' package installed. JFM.cls can take advantage
                 of these fonts,^^Jif you use 'upmath' package.^^J}%
      }
  \else
  \fi
\fi

\ifCUPmtlplainloaded \else
  \checkfont{msam10}
  \iffontfound
    \IfFileExists{amssymb.sty}
      {\typeout{^^JFound AMS Symbol fonts on the system, using the
                'amssymb' package.^^J}%
       \usepackage{amssymb}%

      }{}
  \fi
\fi

\ifCUPmtlplainloaded \else
  \IfFileExists{amsbsy.sty}
    {\typeout{^^JFound the 'amsbsy' package on the system, using it.^^J}%
     \usepackage{amsbsy}}
    {\providecommand\boldsymbol[1]{\mbox{\boldmath $##1$}}}
\fi

\newcommand\hatR{\skew3\hat{R}}      

\newsavebox{\astrutbox}
\sbox{\astrutbox}{\rule[-5pt]{0pt}{20pt}}

\newcommand\shalf{\ensuremath{{\scriptstyle\frac{1}{2}}}}

\title[Drag and lift forces on a counter-rotating cylinder]
{Drag and lift forces on a counter-rotating cylinder in rotating
flow}

\author[C. Sun, T. Mullin, van Wijngaarden and D. Lohse]%
{Chao Sun$^1$, Tom Mullin$^2$, Leen van Wijngaarden$^1$, and Detlef
Lohse$^1$}

\affiliation{$^1$ Physics of Fluids Group, Faculty of Science and
Technology, J. M. Burgers Centre for Fluid Dynamics, MESA+ and
Impact Institutes, University of Twente, The Netherlands.
\\[\affilskip] $^2$
Manchester Centre for Nonlinear Dynamics, University of Manchester,
Oxford Road, Manchester M13 9PL, United Kingdom.}

\date{\today}

\begin{document}

\maketitle

\begin{abstract}

Results are reported of an experimental investigation into the
motion of a heavy cylinder free to move inside a
water-filled drum rotating around a horizontal axis. The cylinder is
observed to either co- or, counter intuitively, counter-rotate with
respect to the rotating drum. The flow was measured with particle
image velocimetry (PIV), and it was found that the inner cylinder significantly altered
the bulk flow field from the solid-body rotation found for a
fluid filled drum. In the counter-rotation
case, the generated lift force allowed the cylinder to freely rotate
without contact with the drum wall. Drag and lift coefficients
of the freely counter-rotating cylinder were measured over a wide
range of Reynolds numbers, 2,500 $<$ Re $<$ 25,000, dimensionless
rotation rates, 0.0$ <  \alpha < $1.2, and gap to cylinder diameter
ratios 0.003 $< G/2a <$ 0.5. Drag coefficients were consistent with previous measurements on a cylinder in a uniform
flow. However, for the lift coefficient considerable larger values
were observed in the present measurements. We found the
enhancement of the lift force to be mainly caused by the vicinity of the wall.

\end{abstract}

\section{Introduction}


The flow around a rotating cylinder is both of fundamental interest
and  of importance in many practical applications, such as flow
control \cite[]{tok91jfm, tok93jfm,mit03jam}  and the motion of
submersed bodies \cite[]{davis}. There have been a number of
investigations into the drag and lift forces which act on a rotating
cylinder in a uniform flow. For example, \cite{bad90jfm} report the
results of a numerical and experimental study of the influence of
the rotation of a cylinder on the flow structure in the wake over
Reynolds number range $10^3$ to $10^4$ and show significant
suppression effects. (The Reynolds number $Re$ is defined here as
$Re= \frac{u_{fs}d}{\nu}$ where $u_{fs}$ is the free stream
velocity, $d$ the diameter of the cylinder and $\nu$ is the
kinematic viscosity of the fluid.) \cite{tok91jfm, tok93jfm}
demonstrate in a series of experiments, that both the wake structure
and drag and lift forces can be controlled by modifying the rotation
of a cylinder in a uniform flow. The suppression of vortex shedding
by rotation of a cylinder is also reported by \cite{mit03jfm}  who
perform numerical simulations at the relatively low value of
$Re=200$. \cite{cliffe04} perform a numerical bifurcation study of
the onset of periodic shedding in a channel and find an exchange
between Hopf and pitchfork bifurcations as a function of control
parameters. \cite{tak04pse} measured lift and drag coefficients on a
rotating cylinder in a flow of $Re$ from 0.4$\times$10$^5$ to
1.8$\times$10$^5$. Their results showed that both lift and drag
coefficients depend on the dimensionless rotation rate of the
cylinder for a fixed $Re$. \cite{lab07jfe} measured the separation
points on a rotating circular cylinder in cross flow at $Re$ ranging
from 8,500 to 34,000, focusing on the dramatic effect of the exact
position of the separation points on the experienced forces. One
conclusion which can be drawn from all of these investigations is
that large rotation rates can generate significant lift coefficients
through the Magnus effect although this is at the cost of the power
requirement for rotating the cylinder which increases rapidly with
the rotation rate \cite[]{mit03jfm}.


The presence of a wall near a cylinder may significantly change the
motion of the object  and the forces acting on it. \cite{hu95tcfd}
studied the two-dimensional motion of a freely rotating cylinder in
a viscous fluid between parallel walls of a vertical channel, for
$Re$ in the range from 0.01 to 102. He found that when the cylinder
moves very close to the channel wall, it  rotates in a direction
opposite to that of rolling along the wall in contact. When the
cylinder is far from the wall, its rotation depends on the value of
$Re$. \cite{bea78jfm} experimentally measured the velocity field and
the distribution of mean pressure around a cylinder near a plane
boundary at a value of $Re$ = 4.5 $\times$ 10$^4$, and found that
regular vortex shedding is suppressed, once the gap between the
cylinder and the wall is less than about 0.3 cylinder diameters.
\cite{sum03} and \cite{cao08} studied the forces on a circular cylinder in uniform shear flow, 
and found that there is a lift force pointing from the high velocity side towards the low velocity side
 due to the asymmetrical distribution of pressure around the cylinder. 
\cite{nis07pof} measured the drag coefficients as a function of the
gap to diameter ratio for a non-rotating cylinder at $Re$ values of
0.4 $\times$ 10$^5$ and 1.0 $\times$ 10$^5$. They generated a uniform flow in the wall region by using a
moving ground for eliminating the boundary layer effect. 
They found that the drag coefficient gradually decreases as the
gap ratio increases, but the dependence is very weak. However, the
lift coefficient rapidly increases as the gap to diameter ratio
decreases to less than about 0.5 \cite[]{nis07pof}.

At \textit{low Reynolds numbers},  \cite{ash05prl,yan06jfm} perform
experiments with heavy spheres in a rotating cylinder. They observe
a cavitation bubble in the lubrication layer between sphere and
wall. The presence of a cavitation bubble results in a normal force
which balances the gravitational component acting on the sphere
\cite[]{pro04fd,ash05prl,yan06jfm}. \cite{sed06pof} experimentally
studied the motion of a heavy cylinder in a rotating cylindrical
flow inside a drum in the Stokes flow regime. They observe that the
cylinder rotates very slowly either with or against the direction of
drum. Again, the cavitation bubbles result in a normal force which
balances the effects of gravity of the cylinder. These results are
in accord with the theoretical predictions of \cite{gjeffrey} (see
\cite{jef81}) who predicts zero rotation of an infinite cylinder
translating adjacent to a plane wall in the Stokes limit.

\textit{Large Reynolds number} The aim of the present work is to
study the flow characteristics and forces exerted on a cylinder
which is rotating freely adjacent to the wall of a rotating
fluid-filled drum since this provides a well-defined flow field
around the cylinder. The same rotating drum setup has been used by
several authors \cite[]{nac92, loh03jcp,nie07jfm, blu08jfm,
blu10, blu09preprint2}  to determine the lift and drag
forces on a light particle or bubble, from its equilibrium position.


Our experiments were carried out in the $Re$ range between 2,500 and
25,000. In the experimental work discussed above, the force
measurements on the rotating cylinder at high $Re$ were performed on
a fixed cylinder, and the rotating rate of the cylinder was
controlled externally. In the present experiments, the cylinder
moved freely in the flow and its rotation rate  was controlled by
the flow. The cylinder either co- or counter-rotated depending on
the rotating frequency of the drum and the drag and lift forces were
determined from the balance of forces acting on the cylinder.

The outline of the paper is as follows. The experimental setup is
introduced in Sec. 2. An overview of  the rotation rate and sense of
rotation of the cylinder are reported in Sec. 3. In Sec. 4, the
results of the PIV measurements in the rotating drum are presented,
first for the local velocity field around the cylinder, then for the
flow field in the whole drum in the absence of the cylinder and
subsequently with the cylinder inserted. The results for the drag
and lift coefficients are presented in Sec. 5 and some conclusions
are drawn in Sec. 6.

\section{Experiment}

\begin{figure}
\centering
\includegraphics[trim=0cm 0cm 0cm 0cm,width=0.75\textwidth]{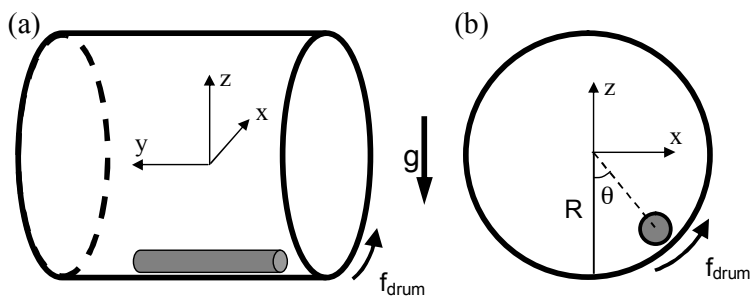}
\caption{A sketch of the experimental setup. (a) Front view and (b)
axial view of the apparatus. The direction of gravity is indicated
by the downward arrow.} \label{fig:setup}
\end{figure}

Schematic diagrams of front and axial views of the experimental
apparatus are given in Fig. \ref{fig:setup}. The plexiglass drum of
length 470 mm and inner radius R = 235 mm was mounted horizontally
and leveled with a precision of less than 0.2 degree. Six solid
PVC (Polyvinyl chloride) cylinders (made from commercially available
extruded PVC cylinders) with density 1,400 kg/m$^3$ were used in the
experiments. Each was 240 mm long and they were of radii $a$ = 7.75,
12.75, 15.5, 20.0, and 30.0 mm respectively. The cylinders were free
to move within the drum, which was filled with de-ionized water. A
motor with a feedback loop control was used to drive the
drum with an accuracy in the rotation frequency of better than 0.01
Hz. The operating frequency, $f_{drum}$, was different for each of
the cylinders and typically lay in the range 0.10 Hz to 0.90 Hz.

A paint mark was put on each of the cylinders, as shown in
fig. \ref{fig:freq} (a, b), in order to measure the angular velocity
of the cylinder. The measurements were conducted once the system
reached an equilibrium state and this typically took 5 to 10 minutes
after each change in $f_{drum}$. The rotation of the cylinder was
filmed with a high speed camera at a typical frame rate of 250
f.p.s. In general, several periods of the rotation of the cylinder
for each f$_{drum}$ were recorded and the rotation frequency of the
cylinder was averaged over the interval.

The azimuthal position of the cylinder in the drum was estimated by
firstly aligning the camera with the axis of the drum. The azimuthal
angle $\theta$ was then calculated based on the measured position of
the cylinder as illustrated in fig. \ref{fig:angle} (a). In order to
measure the thickness of the gap between the cylinder and the wall
of the drum, the camera was mounted horizontally very close to the
cylinder in order to achieve good spatial resolution. Representative
images illustrating this procedure are shown in fig.
\ref{fig:gap_d60mm} (a, b).

\section{Transition between
co-rotation and counter-rotation, regimes of co-rotation and counter-rotation}

\subsection{Transition between co-rotation and counter-rotation}

The cylinder was heavier than the water and sat
at the bottom of the drum when it was at rest. When the drum was set
into motion at a prescribed rotation rate, the cylinder adapted its
rotation frequency and azimuthal position according to the
prescribed value and history of the setting of the drum frequency.
The general behavior of the cylinder was as follows. At lower
$f_{drum}$ values, it intermittently touched the wall, and rotated
in the same direction as the wall of drum. The surface speed of the
cylinder was close to that of the wall, i.e. the significant slip
which is typical of very viscous flows (\cite{sed06pof}) was not
observed here. Beyond a certain threshold frequency, the cylinder
lifted from the wall and rotated in a direction which was
opposite to that of the drum. The rotation frequency $f_{cy}$, the
azimuthal position $\theta$ and the gap width $G$ between the
cylinder and the wall all depended on the drum frequency.

\begin{figure}
\centering
\includegraphics[trim=-0cm 0cm -0cm 0cm,width=1\textwidth]{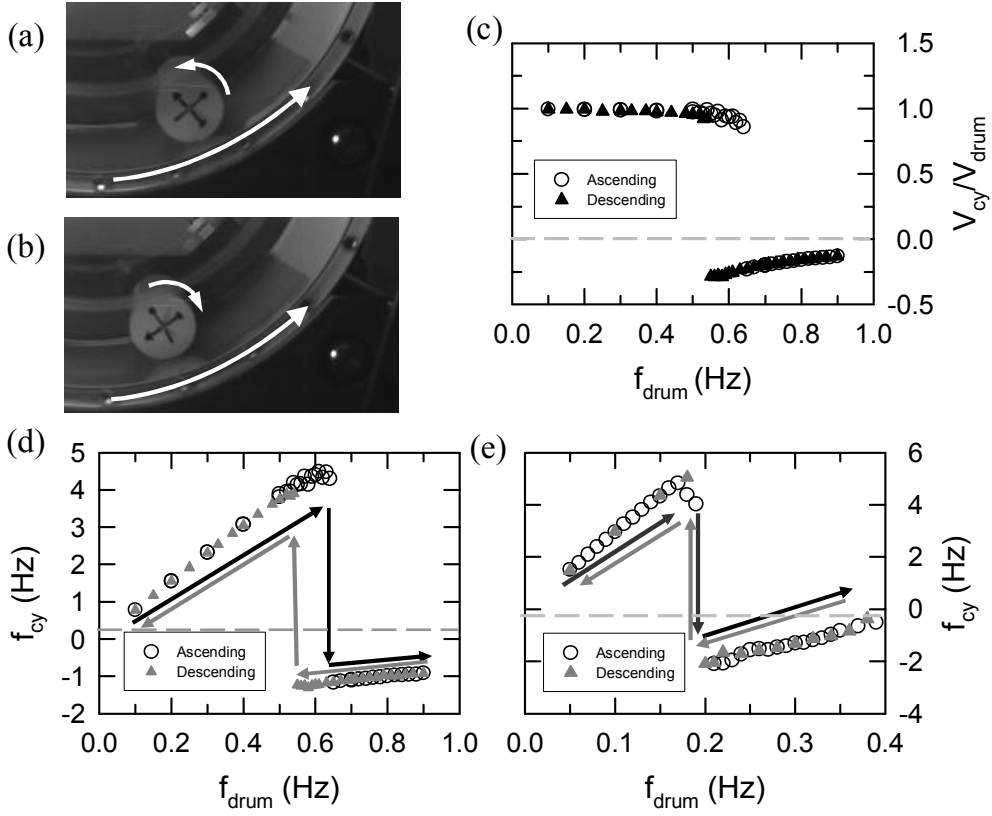}
\caption{Representative frames for the (a) co-rotating (f$_{drum}$ =
0.62 Hz) and (b) counter-rotating (f$_{drum}$ = 0.70 Hz) cylinder ($a$ = 30 mm). Supplementary movies (movie 1 and movie 2) are available for showing the motion of the co- and counter-rotation cylinder. (c) The surface speed ratio between the cylinder ($a$ = 30 mm) and the drum versus rotation frequency of the drum. (d, e) The rotation frequency of the
cylinder, for (d) $a$ = 30 mm and (e) $a$ = 7.75 mm, as function of the rotation frequency of the drum, which was varied both in ascending order (red circles) and
descending order (blue triangles). The gap between the vertical arrows in (d, e) give an
indication of the hysteresis.} \label{fig:freq}
\end{figure}

\begin{figure}
\centering
\includegraphics[trim=-0cm 0cm -0cm 0cm,width=0.8\textwidth]{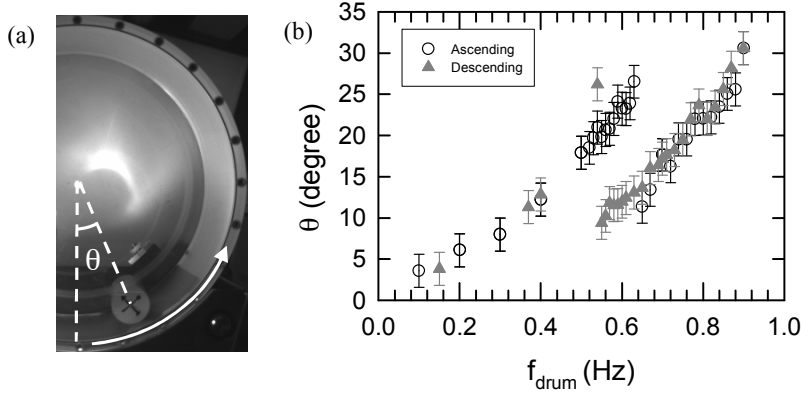}
\caption{(a) Representative frame to demonstrate the measurement of
the azimuthal position. (b) The azimuthal position of the
cylinder ($a$ = 30 mm) as a function of rotation frequency of the drum, which was varied both in ascending order (red circles) and
descending order (blue triangles). The error bars indicate
uncertainty caused by the small wiggling motion of the cylinder, which was found to be less than $\pm$ 2$^{\circ}$. } \label{fig:angle}
\end{figure}

\textit{Cylinder rotation frequency} ~~Plots of the rotation
frequency of the cylinder $f_{cy}$  versus the rotation frequency
of the drum $f_{drum}$, for cylinders of radii 7.75 and 30 mm, are
shown in Fig. \ref{fig:freq} (d, e). The data plotted as open
circles (triangles) were obtained while increasing (decreasing)
$f_{drum}$. The dependence of $f_{cy}$ on $f_{drum}$ for a cylinder
of  radius $a$ = 30 mm is shown in Fig \ref{fig:freq} (d) where can
be seen that the cylinder rotated in same direction as the drum, and
f$_{cy}$ increased with increasing f$_{drum}$ until a critical value
f$_{drum}^{c1}$ = 0.64 Hz was reached. The cylinder suddenly
reversed its rotation direction when f$_{drum}$ exceeded this
threshold. Both the sign and the absolute value of $f_{cy}$ changed
abruptly at this transition. The maximum value of $f_{cy}$ just
before the transition was 4.30 Hz for $f_{drum}$ = 0.64 Hz jumping
to $f_{cy} = -$1.16 Hz for $f_{drum}$ = 0.65 Hz. As shown in Fig.
\ref{fig:freq} (d), beyond the transition the absolute value of
$f_{cy}$ for counter-rotating motion decreases with increasing
$f_{drum}$. The open triangles in the figure represent f$_{cy}$
versus f$_{drum}$ when the experiments were operated with decreasing
f$_{drum}$. The cylinder counter-rotated with $f_{cy} = -$0.91 Hz
when $f_{drum}$ started with 0.90 Hz, and the rotation rate
decreased to $f_{cy} = -$1.23 Hz with decreasing $f_{drum}$ to 0.55
Hz. The cylinder changed to co-rotation when f$_{drum}$ was set to a
smaller value than a second threshold frequency f$_{drum}^{c2}$ =
0.55 Hz, i.e. lower than f$_{drum}^{c1}$ = 0.64 Hz. As shown in the
figure, except in the hysteresis transition region, the curves
$f_{cy}$ for increasing and decreasing $f_{drum}$  lie
on top of each other. The data near the transition frequencies were
measured with increasing or decreasing $f_{drum}$ using the same
increments, and the measurements were performed after a waiting
period of 5 to 10 minutes. Hence, the hysteresis is reported with
confidence.  Plots of the surface speed ratio, defined as
$V_{cy}/V_{drum} = af_{cy}/Rf_{drum}$, of the cylinder with radius
30 mm and the drum are shown in Fig \ref{fig:freq} (c). The surface
speed ratio is very close to 1 for co-rotation, and the absolute
value is smaller than 0.5 for counter-rotation. It can be clearly
seen that the cylinder rotates more slowly in counter- than in
co-rotation with a similar $f_{drum}$.

The observed hysteresis was also found for smaller cylinders, though
the difference between the two thresholds decreased as radius of the
cylinder was reduced. The dependence of $f_{cy}$ on $f_{drum}$ for a
cylinder with diameter 7.75 mm is shown in Fig. \ref{fig:freq} (e).
The overall trend of $f_{cy}$ versus $f_{drum}$ is similar to that
of the cylinder with $a$ = 30 mm. However, the difference between
the two transition frequencies, f$_{drum}^{c1}$ = 0.19 Hz and
f$_{drum}^{c2}$ = 0.18 Hz, for this cylinder is much smaller than
that for the larger cylinder reported above. The observed hysteresis
presumably has its
origins in the history of the flow field which is different when the
drum goes from low to high rotation frequency compared to the
opposite case. We will show later that this is a result of the
interaction of the cylinder with its own wake, which is
qualitatively different for increasing and decreasing $f_{drum}$.
The experimental results also indicate that, as expected, the wake
of a smaller cylinder is weaker than of a larger cylinder, which is
also consistent with the trend for the observed hysteresis.
Hysteresis was also reported by \cite{kan02jsme} in lift force
measurements of a rotating circular cylinder near a moving plane
wall.

\textit{Azimuthal position of the cylinder }~~ Just as the
rotation frequency, the azimuthal position $\theta$ of the
cylinder, shown in Fig. \ref{fig:angle} (a),  is also a function of
the drum rotation frequency f$_{drum}$. For the cylinder with $a$ =
30 mm this dependence is shown in Fig. \ref{fig:angle} (b). The open
circles (triangles) are the measurements made when f$_{drum}$ was
increased (decreased). Hysteresis was also found  in the $\theta$
measurements, and the critical transition frequencies,
f$_{drum}^{c1}$ and  f$_{drum}^{c2}$, are in accord with those from
the frequency measurements discussed above. Apart from the
hysteresis regime, the data sets for increasing and decreasing
$f_{drum}$ can again be collapsed, as shown in fig. \ref{fig:angle}
(b). The angle $\theta$ increases with $f_{drum}$ for both co- and
counter-rotation. However, the angle $\theta$ at which co-rotation
sets in is larger than that for counter-rotation with the same
$f_{drum}$, see fig. \ref{fig:angle} (b). As in the case of the
frequency measurements of $f_{cy}$,  the hysteresis of $\theta$ is reduced for
the smaller cylinder.

\begin{figure}
\centering
\includegraphics[trim=-0cm 0cm -0cm 0cm,width=1\textwidth]{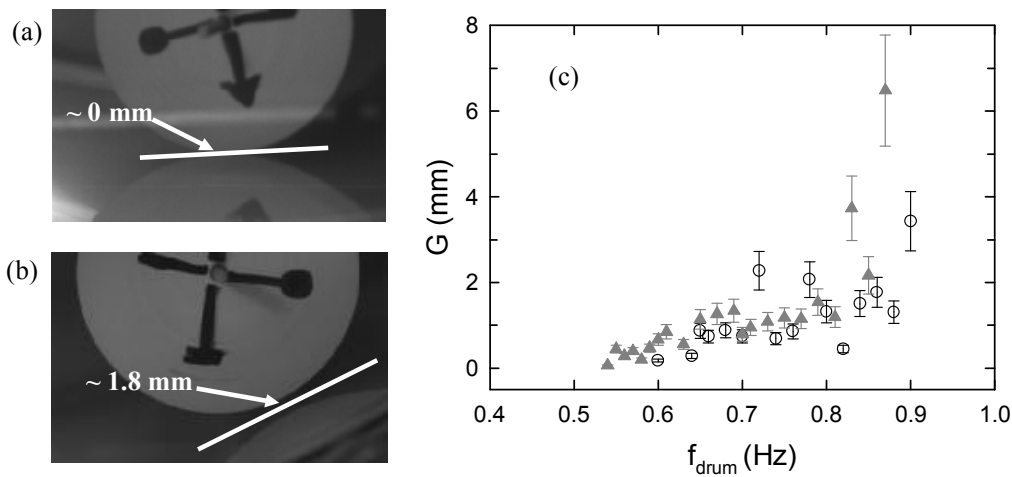}
\caption{The snapshots indicate that (a) the gap thickness for
co-rotation case with $f_{drum}$ = 0.10 Hz; (b) the gap thickness
for counter-rotation with $f_{drum}$ = 0.86 Hz. The radius of the
cylinder is 30 mm, and the solid line marks the wall of the
drum. Its curvature is not seen on the scale of this figure. (c) The measured gap width as a function of
$f_{drum}$ for the cylinder of radius 30 mm in the counter-rotation situation. The drum was operated both in ascending mode (red circles) and
descending mode (blue triangles). The error bar in (c) is 20$\%$ of the gap thickness.}
\label{fig:gap_d60mm}
\end{figure}

\textit{Gap width}~~The cylinder was observed to sit very close to
the drum wall when it was co-rotating with the drum as
shown in fig. \ref{fig:gap_d60mm} (a). In this case, the width of
the gap between the cylinder surface and the drum wall was below the
resolution of the measurements. 
Since the pressure force, a detailed account of which is given in Sec. 5.2, is too small to balance the radial component of gravity, a normal force exerted by the wall must support the cylinder. This is confirmed by Fig. \ref{fig:freq}(c) which shows that the surface speeds of the cylinder are nearly equal to the speed of the drum wall, implying friction dominates. The speed of
the cylinder reduces when the drum is close to the transition
frequency and the cylinder begins to slip with respect to the
wall.

The situation was found to be quite different when the
cylinder counter-rotated with the
drum. Then the cylinder `floated' above the wall instead of being in
contact with it, as can be seen in fig. \ref{fig:gap_d60mm} (b). The
measurements of the gap width $G$ versus $f_{drum}$ for the
counter-rotation cylinder with radius of $a =$ 30 mm are presented
in Fig \ref{fig:gap_d60mm} (c). It can be seen that $G$ is an
increasing function of $f_{drum}$. The hysteretic behavior is the
same for the frequency and angle measurements. The gap width
shown in Fig. \ref{fig:gap_d60mm} (c)  was estimated using snapshots
obtained from the respective experimental condition. However, the
cylinder did not always rotate at precisely the same fixed position
during the measurements, and a small wiggling motion produeced an error. Averaging over many
snapshots gave an error estimate of less than $20\%$ in the
measurements of the gap and this is indicated by the error bars
in the figure.

\bigskip

We separately discuss the observed phenomena for the co- and
counter-rotation cases in the following section.

\subsection{Co-rotation}

\cite{pro03fd} studied the motion of a rigid particle rolling down
an inclined plane in a fluid at low values of $Re$. It was found
that the particle rolled under its own weight, and exhibited both
hydrodynamic slip and contact with the wall. \cite{yan06jfm}
measured the speed ratio of a rough sphere in a rotating drum at low
values of $Re$. They found that the speed ratio between the sphere
and the wall was approximately one when the rotation frequency of
the drum was low so that the sphere moved with the drum via the
frictional contact with the wall. At faster drum rotation the sphere
began to slip with respect to the drum wall and smoothly departed
from the contact regime.

\begin{figure}
\centering
\includegraphics[trim=-0cm 0cm -0cm 0cm,width=0.8\textwidth]{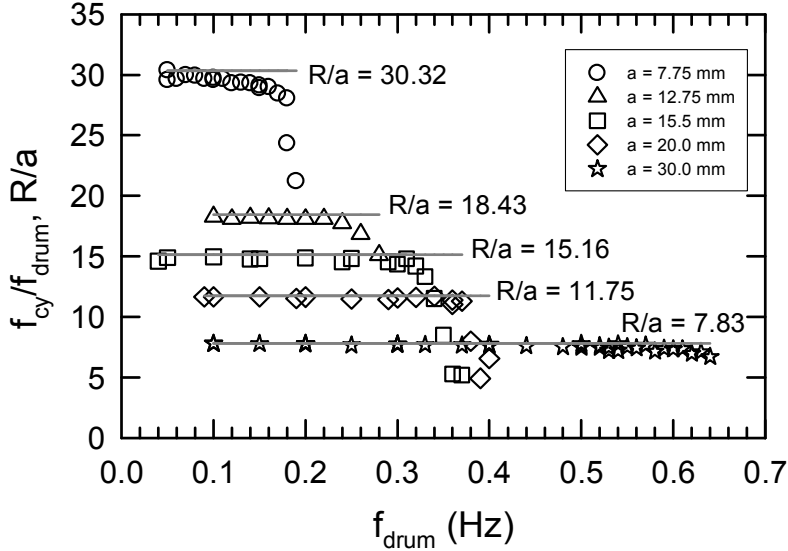}
\caption{Co-rotating case: The ratio between $f_{cy}$ and $f_{drum}$ as a function of $f_{drum}$ for different
 cylinders. The solid lines represent the ratio of the
radii of the drum and the cylinder.}
\label{fig:co_rot}
\end{figure}

Not unexpectedly, frictional interaction between the heavy cylinder
and the drum forces the former to co-rotate. In the limiting case,
the cylinder completely rolls along the drum wall. In this
situation, the rotation frequencies of the cylinder and the drum are
inversely proportional to their radii, i.e. f$_{cy}$/f$_{drum}$ =
R/a. The ratio between two rotation frequencies is plotted as a
function of f$_{drum}$ (in the co-rotation case) for a number of different cylinders in Fig. \ref{fig:co_rot}. The solid lines depict the
ratios between the drum radius and the respective cylinder radii, and hence indicate friction dominates the
motion for the different cylinders. Fig. \ref{fig:co_rot} shows that the measured data collapse on these
lines, and this indicates that the cylinder indeed moved with the drum (via
the frictional contact with the wall). The deviation between the data
and the line at higher $f_{drum}$ suggests that the cylinder then
started to slip. The cylinder
frequency was much lower than that of the rolling motion when
f$_{drum}$ was close to the critical frequency, where the transition to counter-rotation
took place.

\subsection{Counter-rotation}

As discussed above, the cylinder counter-rotated once the drum
rotation frequency was larger than a threshold value. In
counter-rotation, the cylinder freely rotated without contact with
the wall and there was a significant gap between it and
the wall. Here we focus on the cylinder rotation frequency, the
azimuthal location, and the gap width in the counter-rotation case.

\begin{figure}
\centering
\includegraphics[trim=-0cm 0cm -0cm 0cm,width=1\textwidth]{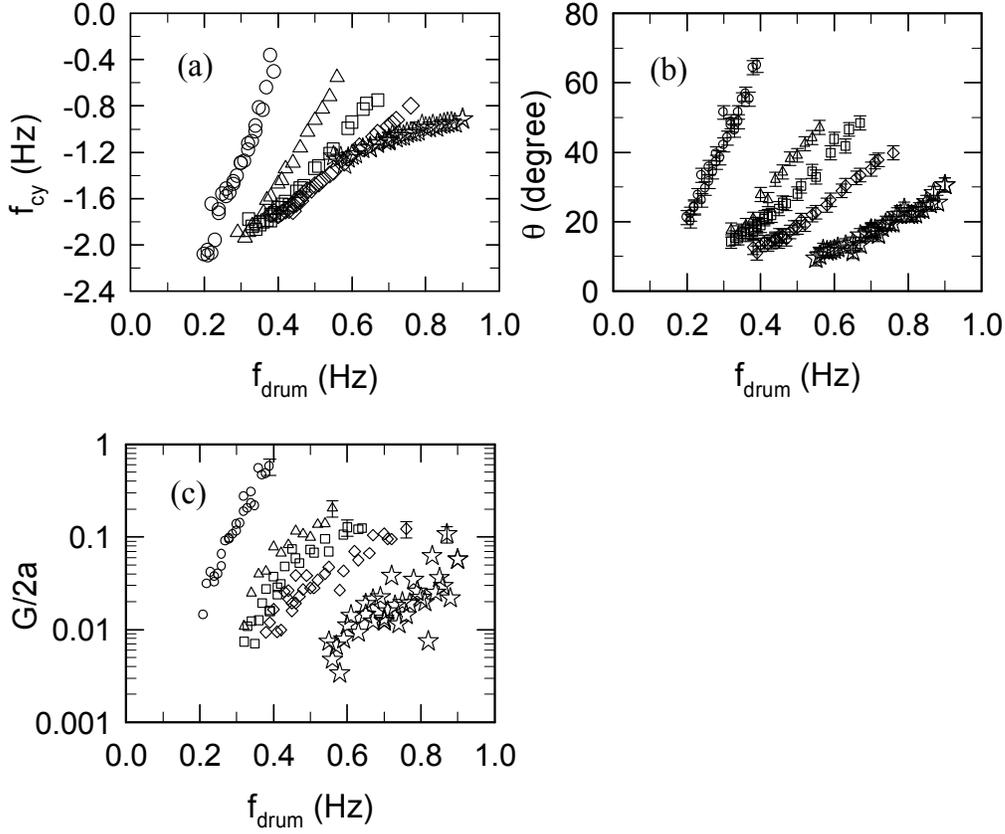}
\caption{Counter-rotating case: The measured (a) cylinder rotation frequency (b)
azimuthal position of the cylinder,  and (c) the normalized gap width between the cylinder and the drum wall as a function of $f_{drum}$
for the cylinders of different radii. The measurement error bar
shown in (b) is $\pm$ 2$^\circ$. The error bar in (c) is 20$\%$ of the gap thickness, only maximum error bar for each cylinder is shown. The symbols in the plots are same with Fig. \ref{fig:co_rot}; Open circles: $a$ = 7.75 mm, open triangles: $a$ = 12.75 mm, open squares: $a$ = 15.5 mm, open diamonds: $a$ = 20.0 mm, open stars: $a$ = 30.0 mm. } \label{fig:counter_rot}
\end{figure}

The counter-rotating cylinder self-selected a rotation frequency and
an azimuthal location according to the set frequency of the drum and
the radius of the cylinder. The relationship between f$_{cy}$ and
f$_{drum}$ for  different cylinders in counter-rotation is shown in
Fig. \ref{fig:counter_rot} (a). The cylinder counter-rotated with
respect to the drum, so that f$_{cy}$ in the plot is negative. The
absolute value of f$_{cy}$  decreased approximately linearly with
increasing f$_{drum}$ for all cylinders, the cylinder tended to slow down 
as the speed of the drum increases, i.e. the opposite to the case of
co-rotation.  As shown in Fig.
\ref{fig:counter_rot} (a), the cylinder with  $a$ = 7.75 mm began to
counter-rotate at 2.1 Hz when f$_{drum}$ = 0.2 Hz, and slowed down
to 0.4 Hz when f$_{drum}$ was 0.4 Hz. Above this frequency the
cylinder started to wiggle around instead of rotating steadily at a
fixed position.
\cite{ste06, ste10} studied the wake of a rolling cylinder/sphere along a wall, and they found that 
the wake becomes unsteady for Reynolds number above a few hundred. In the present studies, the cylinders 
freely counter-rotated with the drum. We also found that the cylinder motion was no longer stable
when the drum frequency was increased further. The transition Reynolds number, from stable rotation to unstable motion, depended on the cylinder radius. Here we only report measurements for stable
rotation of the cylinder.

The frequency of $f_{drum}$ at which counter-rotation was proportional to the diameter of
the cylinder and ranged ranges from 0.2 Hz for $a$ =
7.75 mm to 0.55 Hz for $a$ = 30 mm. Each cylinder lost stability to
time-dependent motion and started to wiggle when f$_{drum}$ was
larger than the maximum value of $f_{drum}$ given on the plot. This limiting frequency for stable rotary
motion of the cylinder is also proportional to its diameter as
can be seen in Fig. \ref{fig:counter_rot} (a). It
changes from 0.4 Hz for the cylinder of $a$ = 7.5 mm to 0.9 Hz for
the $a$ = 30 mm cylinder.

The self-selected azimuthal position $\theta$ of the cylinder also
depends on f$_{drum}$ as shown in Fig. \ref{fig:counter_rot} (b).
The cylinder generally rotated stably during the measurements,
however, it was never perfectly still and there was always a small
amplitude wiggling motion present. This was analyzed in detail for a
particular case and the error induced by the wiggling motion was
found to be less than 2$^\circ$ as indicated by the error bars
marked on the plot. 
The angle ($\theta$)
at which counter-rotation starts increases approximately
linearly with f$_{drum}$ for all cylinders, but has a different slop for each.

The gap between the cylinder and the drum wall was also
self-selected by the cylinder for given $f_{drum}$.  The measured
gap width is shown in Fig. \ref{fig:counter_rot} (c) plotted as
a function of $f_{drum}$ for each individual cylinder when in
counter-rotating. The data clearly indicate that the gap width is an
increasing function of f$_{drum}$ and for fixed $f_{drum}$, a larger
gap was found for smaller cylinders. It is quite remarkable that the
gap is of the order of a millimeter, which is significantly larger than the
micrometer range found by \cite{sed06pof} in the Stokes flow regime.
The maximum gap width in this case was several millimeters and this allowed us
to visualize the flow around the cylinder using particle image
velocimetry (PIV).

The dependence of the rotation frequency $f_{cy}$ on its
azimuthal location $\theta$ and the gap thickness $G$ presented in
this section will all be used to derive the drag and lift
coefficients discussed in Sec. 5.

\section{Velocity measurement}

The flow field around the cylinder determines the forces exerted on it. Particle Image Velocimetry (PIV) was employed to study
the flow around the cylinder in the $x-z$ plane, see Fig.
\ref{fig:setup}, for both co- and counter-rotation situations.

\subsection{Visualization of local flow around cylinder}

\begin{figure}
\centering
\includegraphics[trim=-0cm 0cm -0cm 0cm,width=0.6\textwidth]{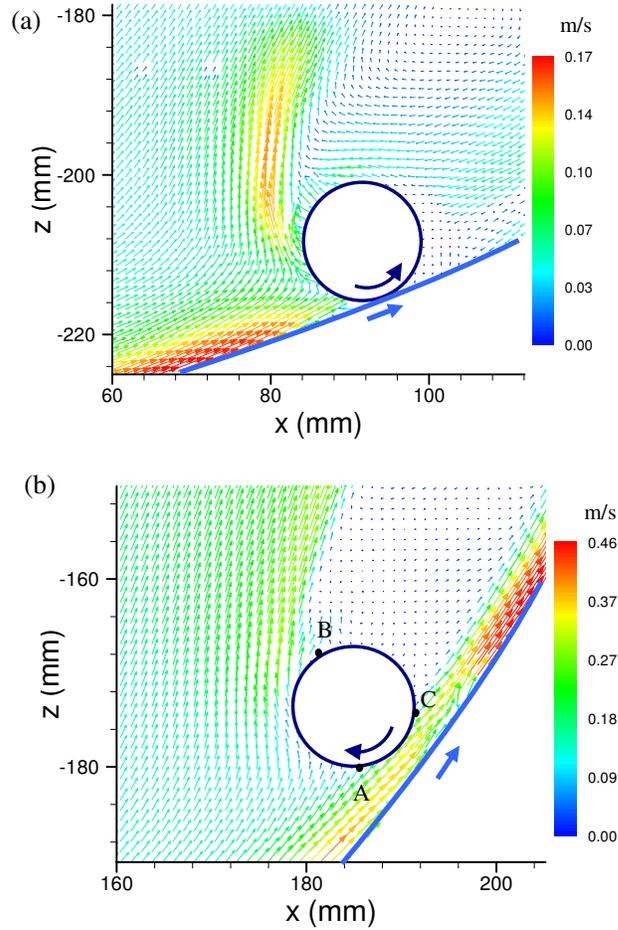}
\caption{The measured time-averaged velocity around the cylinder with radius $a$ = 7.75 mm. The velocity maps were averaged with 300 instantaneous velocity snapshots with a frequency of 50 Hz. The magnitude of the velocity $(U^2+W^2)^{1/2}$ was coded in color scale in units of m/s.  (a) The cylinder co-rotates with the drum with the set frequency $f_{drum}$ = 0.15 Hz. (b) The cylinder counter-rotates with the drum operating at $f_{drum}$ = 0.35 Hz. Supplementary movies (movie 3 and movie 4) show the time evolution of the instantaneous velocity maps. } \label{fig:Flow_pattern}
\end{figure}

The time-averaged velocity map around the co-rotating cylinder with
radius 7.75 mm is shown in Fig. \ref{fig:Flow_pattern} (a). The
rotating frequency of the drum was set to $f_{drum} =$ 0.15 Hz, and
the cylinder was co-rotating with a frequency of 4.33 Hz; the
azimuthal position was $\theta$ = 23.5 $^o$. The ratio between the
two rotation frequencies is $f_{cy}/f_{drum}$ = 28.87, which is
slightly smaller than the radii ratio $R/a$ = 30.32. This suggests
that the cylinder did not completely roll along the wall of the
drum,  and there was some slipping motion between the cylinder and
the wall. As shown in Fig. \ref{fig:Flow_pattern} (a), the flow
cannot pass between the cylinder and the wall and this induces a
strong upward flow. Associated with this upward flow is a shear
stress exerted on the cylinder. With increasing frequency of the
drum the couple on the cylinder become larger and eventually
overcomes the forcing from the wall and the cylinder starts to
rotate in a sense opposite to the drum.

As soon as the cylinder starts to counter-rotate, a gap develops between the cylinder and
the wall.  The time-averaged velocity map
around the cylinder for the counter-rotating case is shown in Fig.
\ref{fig:Flow_pattern} (b). The cylinder was lifted above the wall,
reaching a distance of about 3.3 mm. As shown in Fig.
\ref{fig:Flow_pattern} (b), in this case, the maximum velocity is close to the wall, i.e.
inside the gap and hence the primary route for the mass flux is
through the gap. The small difference is within the measurement
error.
The mass flux coming from upstream must be transferred to downstream through the gap. The flow velocity inside the gap is quite close to the upstream flow velocity, and the small difference may be induced by measurement error. Beyond the separation point, the acceleration is caused by the interaction between dividing streamline and the wake.

When a cylinder moves in a {\it uniform} flow far from a wall and without rotatory motion, the separation points are located {\it symmetrically} with respect to a line joining the forward separation point with the center of the cylinder.  In the present case of  the counter-rotating cylinder, the separation points are {\it not} symmetric.
The boundary layer on the cylinder
cannot be resolved using PIV. However, an estimate of the
location of the stagnation point can be made using the
information on the flow direction adjacent to the cylinder
surface. As shown in Fig. \ref{fig:Flow_pattern} (b),
the separation point (C) is located near the position where high
velocities exist in the gap. The counter-rotation motion and the
vicinity of the wall forces the separation point (B) to move
downstream and the stagnation point (A) to move towards the wall.
Hence the location of the stagnation and  separation points imply
that the tangential shear stress on the surface between the point A
and C is larger than that between A and B. The overall tangential
shear stress on the surface of the cylinder therefore drives the
counter-rotation of the cylinder. The lift force acting on the
counter-rotating cylinder acts to balance forces resulting from the
normal component of gravity and inertial pressure force.  The movement
of the separation points for a cylinder near a wall were also
studied in experiments by \cite{lab07jfe}, who found a similar
trend. As we will show later, a large lift coefficient on the
cylinder can be
obtained and this arises principally because the nearby wall.

\bigskip

The PIV velocity measurements were conducted under different
experimental conditions to quantify the flow field in the drum. The
measurements were performed in the central ($x-z$ plane at $y$ = 0)
as shown in Fig. \ref{fig:setup}. The time-averaged velocity maps
presented in the following section were obtained by averaging 50 to
300 instantaneous velocity frames which were captured with a
measurement frequency of 50 Hz. It was found that 50 velocity frames
are sufficient to achieve a convergent time-averaged velocity map.

\subsection{The velocity field without cylinder}

A check was first performed to establish  whether the flow in the
drum without a cylinder was in solid-body rotation. \cite{blu08jfm}
estimated the spin-up time to achieve solid-body rotation  starting
from rest is approximately 4 minutes for water. All data in the
present measurements were taken after this waiting period.

\begin{figure}
\centering
\includegraphics[trim=-0cm 0cm -0cm 0cm,width=1\textwidth]{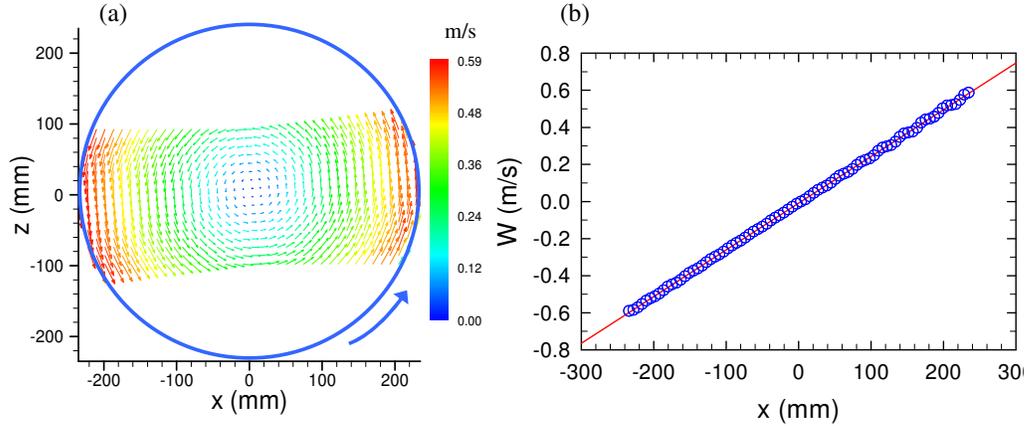}
\caption{(a) The time-averaged velocity vector map, in the drum without a cylinder, measured with the set frequency $f_{drum}$ = 0.40 Hz. For better readability, a coarse-grained map is shown here with only 1/4 of the measurement arrow density.The magnitude of the velocity was coded in color scale in
m/s. The time average was taken over a period of 6 seconds corresponding to 300 velocity frames.
(b) The horizontal profile of the vertical velocity $W$ extracted from the measured velocity maps in (a). The straight line corresponds to the fitting result ($f_{drum}$ = 0.401 Hz), which is very close to solid-body rotation.}
\label{fig:piv-whole.eps}
\end{figure}

As a check, it
was decided to measure the flow field directly. The
time-averaged velocity field measured in the $x-z$ plane of the drum
operated at a rotation frequency of $f_{drum}$ = 0.40 Hz is shown
in Fig. \ref{fig:piv-whole.eps} (a). Practical limitations of the
laser light intensity meant that only the central part of the
vertical direction was measured. The selected measurement area was
470 mm $\times$ 196 mm, which corresponds to $-$R to R in $x$
direction and $-$0.42R to 0.42 R in $z$ direction. The spatial
resolution between two vector arrows was 5.93 mm in both $x$ and $z$
directions.  A profile of the vertical velocity along
the $x$-axis, extracted from the measured velocity map in $z = 0$,
is plotted in figure \ref{fig:piv-whole.eps} (b). The vertical
velocity depends linearly on the horizontal position $x$, 
confirming solid-body rotation. The rotation
frequency for the drum based on the measured slope is 0.401 Hz,
which is very close to the set value of 0.40 Hz.

\begin{figure}
\centering
\includegraphics[trim=-0cm 0cm -0cm 0cm,width=1\textwidth]{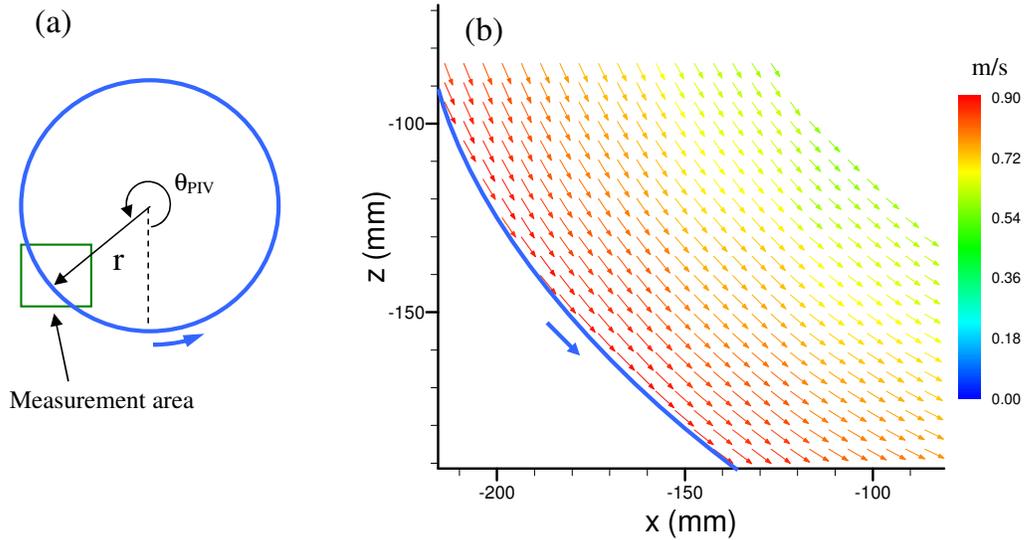}
\caption{(a) A sketch of the measurement area for the high resolution PIV measurement. The angular location of the measurement center is $\theta_{PIV} $ = 256.8$^{\circ}$. (b) The time-averaged velocity vector map measured at $f_{drum}$ =  0.60 Hz. For better readability, a coarse-grained map is shown here with only 1/9 of the measurement arrow density.
The magnitude of the velocity was coded in color scale in 
m/s. The time average was taken over a period of 6 seconds, corresponding to 300 velocity frames.  }
\label{fig:piv-whole-hr.eps}
\end{figure}

To study the flow
velocity near the wall of the drum in detail, higher resolution PIV
measurements were performed in this region for frequencies from 0.10
to 0.90 Hz. The measurement area was 134 mm $\times$ 107 mm, which
offered us a spatial resolution 1.7 mm between two vector arrows. A
sketch of the measurement location is shown in Fig.
\ref{fig:piv-whole-hr.eps} (a) where the angular location of the
measurement center was $\theta_{PIV} $ = 256.8$^{\circ}$. The
time-averaged velocity vector map measured with a drum rotating
frequency of 0.60 Hz is shown in  Fig. \ref{fig:piv-whole-hr.eps}
(b). A coarse-grained velocity map is reproduced here with only 1/9
of the measurement arrow density to enable a printable figure. The
magnitude of the velocity was coded in color scale
in units of m/s. In fact, the magnitude of the velocity was found to be almost
identical with the value of the tangential velocity $V_{\theta}$ because the flow
is in almost perfect solid-body rotation. The standard derivations for horizontal and vertical velocity 
components, averaged in the whole map, were found to be only 0.025 m/s, which is far smaller than the mean flow velocity. This suggests that the flow is stationary.

\subsection{The velocity field with an inserted cylinder in the drum}

Solid-body rotation was disturbed when a cylinder was placed in the
drum. We will now quantify this effect through PIV measurements and
will, in particular, clarify whether the flow field returns to
solid-body rotation after traveling a certain distance downstream
the cylinder.

\begin{figure}
\centering
\includegraphics[trim=-0cm 0cm -0cm 0cm,width=0.9\textwidth]{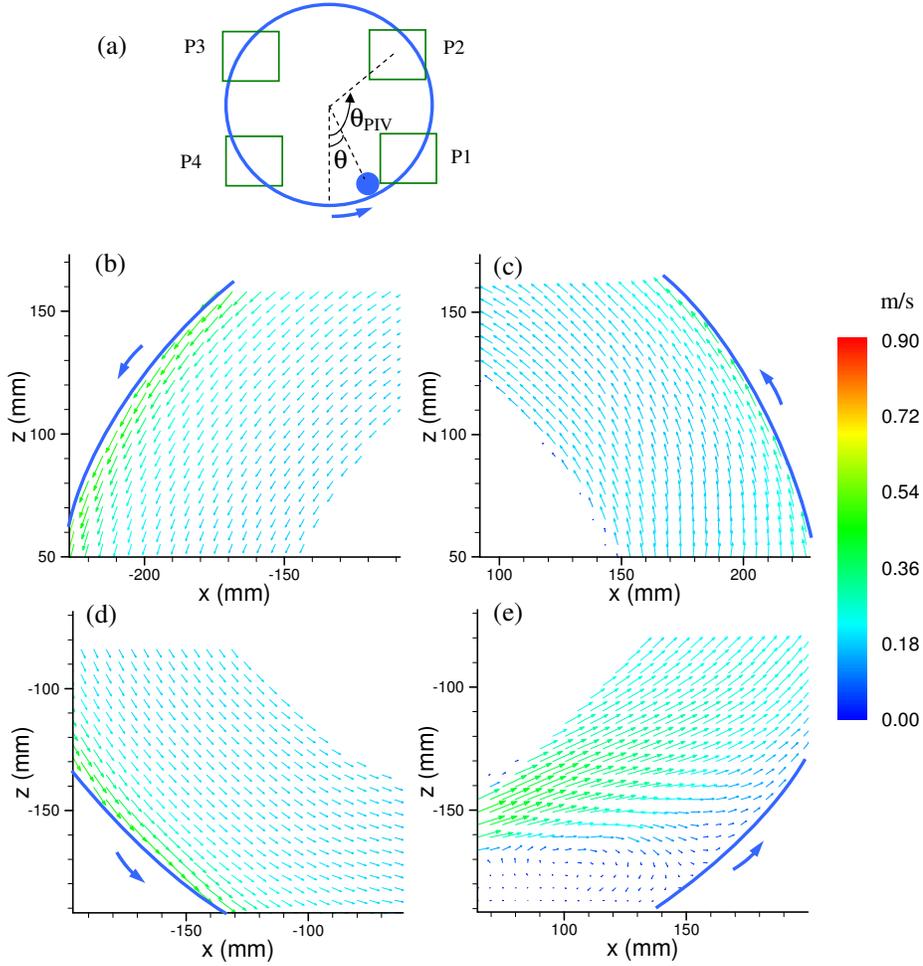}
\caption{(a) A sketch of the cylinder location and the PIV measurement positions in the drum. The radius of the cylinder was 30 mm in this measurement and the set drum rotating frequency was $f_{drum}$ = 0.60 Hz. (b-e) The time-averaged velocity maps measured at P3, P2, P4 and P1.  For the sake of clarity, the coarse-grained maps are shown here with only 1/9 of the measurement arrow density. The magnitude of the velocity was coded with the same color scale as that in Fig. \ref{fig:piv-whole-hr.eps} (b) with the same set frequency.  The time average was taken over a period of 1 seconds corresponding to 50 velocity frames. }
\label{fig:a30mm_4profiles}
\end{figure}

Firstly, the evolution of the flow velocity behind the cylinder was
studied by measuring the velocity distributions at different
azimuthal positions. A sketch of the cylinder location and the PIV
measurement positions in the drum is shown in Fig.
\ref{fig:a30mm_4profiles} (a). The radius of the cylinder was 30 mm
in this measurement. The cylinder counter-rotated with $f_{cy} =
-$4.4 Hz when the drum rotation frequency was set to $f_{drum}$ =
0.60 Hz. The gap between the cylinder and the wall was approximately
0.2 mm, and the respective azimuthal location of the cylinder center
was $\theta$ = 12 $\pm$ 2 $^\circ$.  High resolution PIV
measurements were performed to measure the flow field at different
distances from the cylinder. These were made at four different
azimuthal positions with $\theta_{PIV} $ = 48$^\circ$ (P1),
122$^\circ$ (P2), 239$^\circ$ (P3), and 312$^\circ$ (P4),
respectively. The respective distances traveled by the wake
normalized by the cylinder diameter, i.e. $(\theta_{PIV}-\theta)
(R-a)/ 2a$, were 2.1, 6.5, 13.5, and 17.9, with $\theta$ expressed
in radians.

The time-averaged velocity maps at the measurement areas P1, P2, P3,
and P4 are presented in Figures \ref{fig:a30mm_4profiles} (e, c, b,
d), respectively. The color code for the velocity
is the same as that in Fig. \ref{fig:piv-whole-hr.eps} (b), since
the experiments were carried out at the same drum rotation
frequency. At  $f_{drum}$ = 0.60 Hz the maximum velocity is 0.90 m/s
for a drum without a cylinder, as shown in Fig.
\ref{fig:piv-whole-hr.eps} (b). However, the maximum flow velocity,
at the same $f_{drum}$, for all four positions is only around 0.4
m/s, which is significantly smaller. The measurement area P1
measures the velocity field just downstream the cylinder, the wake
generated by the cylinder is clearly visible in Fig.
\ref{fig:a30mm_4profiles} (e). The flow certainly develops with
increasing distance from the cylinder. The vortex disappears in the
measurement area P2, and some vectors tend to align in tangential
direction. However, a significant number of vectors do not point in
the tangential direction, especially in the lower part, as shown in
Fig. \ref{fig:a30mm_4profiles} (c). The flow continues to develop
with increasing distance from the cylinder. As shown in
\ref{fig:a30mm_4profiles} (b, d), almost all velocity vectors in the
velocity maps measured at P3 and P4  point in tangential direction.
Moreover, the velocity patterns in measurement areas P3 and P4 are
very similar as shown in Fig. \ref{fig:a30mm_4profiles} (b, d). This
suggests that the flow has achieved a steady state at the
measurement area P3 and P4. However, it is far from the solid-body
rotation without a cylinder, as shown by the clear difference of the
velocity patterns and the velocity magnitudes between Fig.
\ref{fig:piv-whole-hr.eps} (b) and Fig. \ref{fig:a30mm_4profiles}
(d).

\begin{figure}
\centering
\includegraphics[trim=-0cm 0cm -0cm 0cm,width=0.7\textwidth]{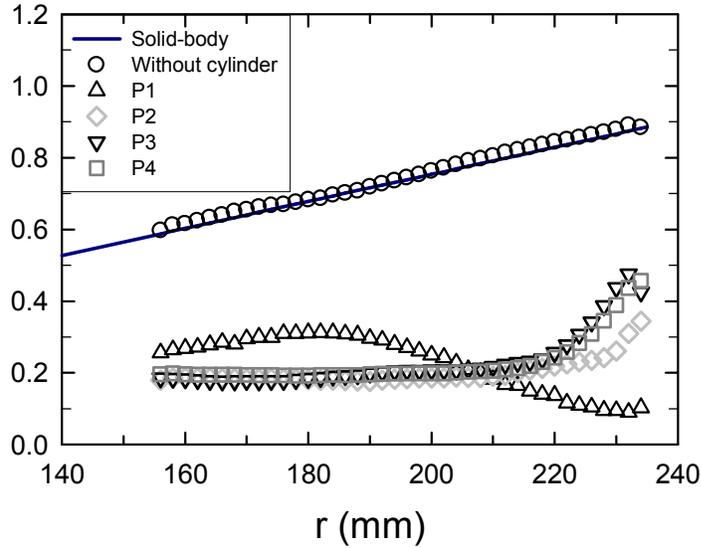}
\caption{The tangential velocity $V_{\theta}(r)$ versus the radial distance at different measurement positions. The drum rotation frequency is $f_{drum} = $ 0.60 Hz and $a$ = 30 mm. The respective velocity profile, with the same $f_{drum}$, in the drum without cylinder is plotted with open circles and that for a solid-body rotation is shown as solid line. A considerable velocity reduction is seen throughout.}
\label{fig:a30mm_v_cut}
\end{figure}

In order to compare the velocity profiles quantitatively, the
tangential velocity $V_{\theta}(r)$ plotted as a function of the
radial distance from the drum center (r) at each measurement
position is shown in Fig. \ref{fig:a30mm_v_cut}. The radial profile
was obtained by averaging the velocity vectors with the same $r$, with
angle between $\theta_{PIV}-2.5^{\circ}$ and
$\theta_{PIV}+2.5^{\circ}$ in each of the measurement areas. For
comparison, the velocity profile in the drum without a cylinder,
extracted from Fig. \ref{fig:piv-whole-hr.eps} (b), is also shown
plotted in the figure using open circles. The profile for solid-body
rotation with the same $f_{drum}$ is also shown in the plot (solid
line). As shown in Fig. \ref{fig:a30mm_v_cut}, again the velocity
profile in the drum without cylinder agrees with the solid-body
rotation. However, all the profiles, with the cylinder at four
positions, do not follow solid-body rotation in either the magnitude
of the velocities or the shape of the profiles. The velocity profile
at P1 is clearly disturbed significantly by the cylinder.
It develops with distance from the cylinder, and the velocity
profiles at P3 and P4 are almost identical. This suggests that the
flow at P3 (and P4) is already in a stationary stable state, since
it does not change with increasing distance from the
cylinder. This well developed velocity profile at measurement
position P4 corresponds to the incoming velocity profile for the
cylinder. The profile is not linear, as it would be for solid-body
rotation, but is instead curved. The velocity rapidly decreases at a
short distance from the wall, and it changes slowly with further
increase of the distance from the wall, as shown in Fig.
\ref{fig:a30mm_v_cut}.

\subsection{The incoming velocity}

\begin{figure}
\centering
\includegraphics[trim=-0cm 0cm -0cm 0cm,width=0.7\textwidth]{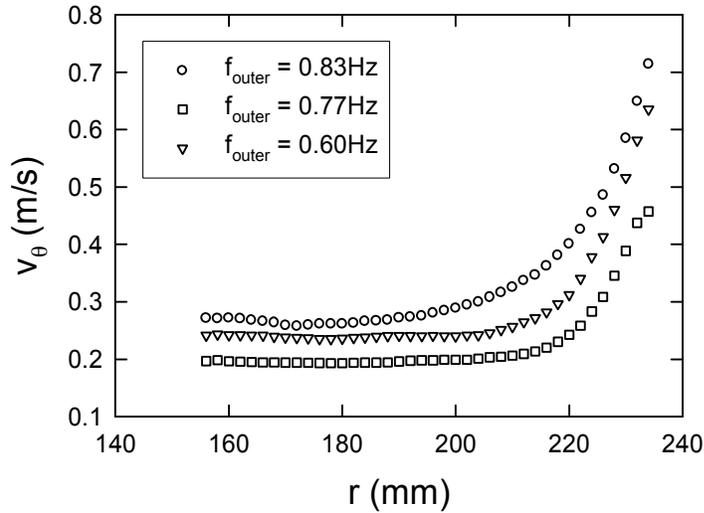}
\caption{The tangential velocity profiles versus the radial distance measured at the measurement position 4 for various set frequencies. The cylinder inside the drum has an radius of 30 mm. }
\label{fig:profile_p4_a30}
\end{figure}

\begin{figure}
\centering
\includegraphics[trim=-0cm 0cm -0cm 0cm,width=0.7\textwidth]{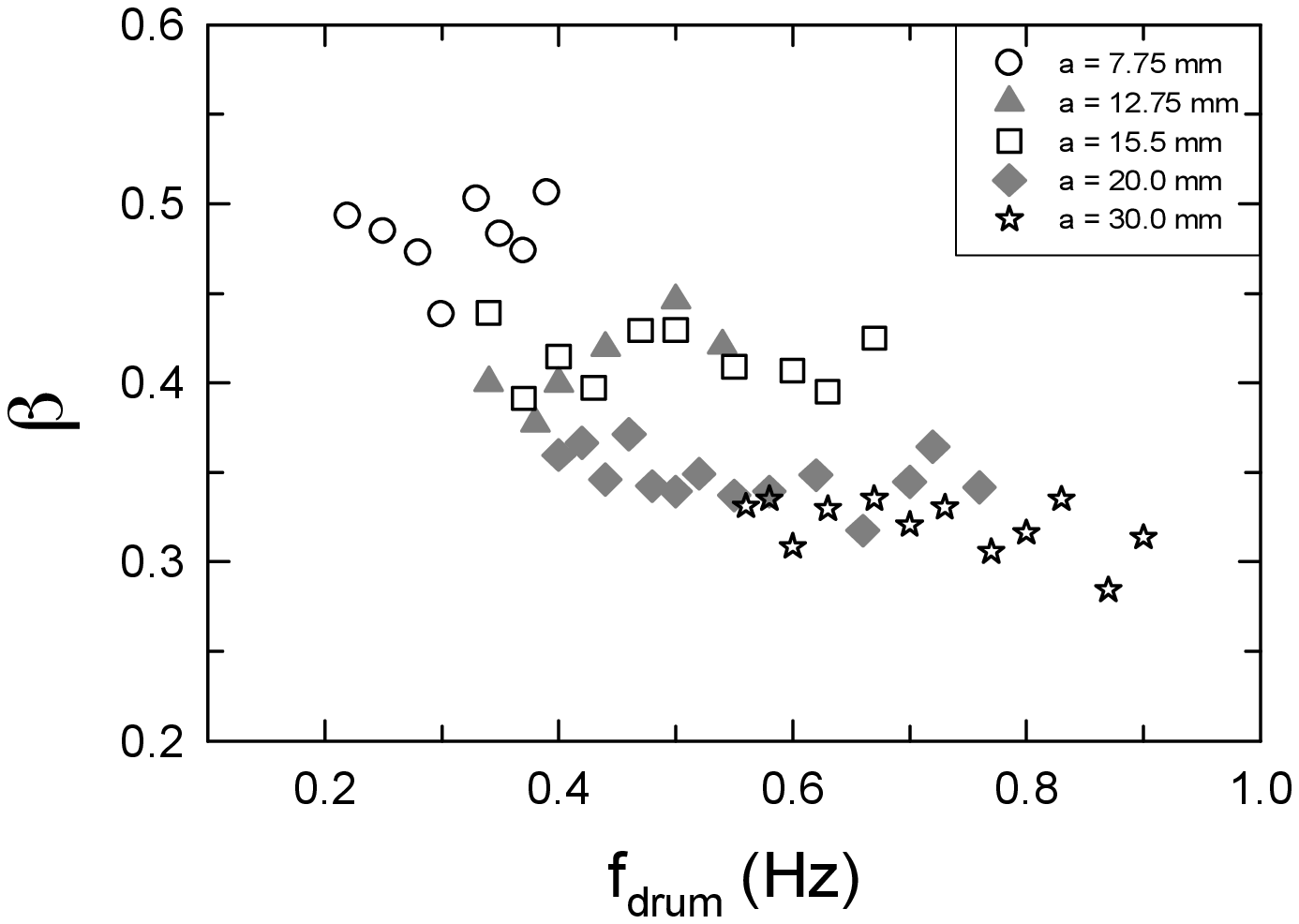}
\caption{The normalized incoming velocity plotted
as a function of the drum frequency for different diameter
cylinders. }
\label{fig:incoming_vel}
\end{figure}

As discussed above, the incoming velocity $V_{0}$ to the cylinder
does not correspond to solid-body rotation. In order to quantify the
effect of this, velocity fields were measured at the position P4 for
various cylinders at different drum frequencies. Since we focus on
the force measurements on a counter-rotation cylinder, the
measurements were only performed in the counter-rotation regime. The
results for the tangential velocity are presented in Fig.
\ref{fig:profile_p4_a30}. The cylinder with radius $a =$ 30 mm
counter-rotated with the drum over the frequency range used here.
The profile shapes are similar, and, as expected higher velocities
were found for higher $f_{drum}$. For the various cylinders and
various $f_{drum}$ the radial velocity profiles $V_{\theta}(r)$ at
P4 were determined. Those for $a = $ 30 mm are shown in Fig.
\ref{fig:profile_p4_a30}. As discussed previously, the gap between
the cylinder and the wall of the drum is very small. We neglect this
for the calculations of the incoming velocity.

With measured tangential velocity $V_{\theta}(r)$ and cylinder radius $a$, 
we define the incoming velocity $V_0$  as the
average of the tangential velocity between R and (R-2a),

\begin{equation}
V_{0} = \frac{1}{2a} \displaystyle\int^{R-2a}_{R} V_{\theta}(r)dr.
\end{equation}

The incoming velocity based on a solid-body rotation is $2 \pi
f_{drum} (R-a)$. The ratio between the measured incoming velocity
$V_{0}$ and the one based on a solid-body rotation measures the
degree of the perturbation by the cylinder. This ratio is defined by
\begin{equation}
\beta =\frac {V_{0}} {2 \pi f_{drum} (R-a)}.
\end{equation}

We measured the ratio for every cylinder under several frequencies
in its parameter range, as shown in Fig. \ref{fig:incoming_vel}. The
data indicates that $\beta$ is roughly constant for a given cylinder
radius $a$ in the measured parameter regime and becomes smaller for
larger cylinders. Averaged over various $f_{drum}$ values of the
ratios $\beta$ are $\overline{\beta}(a)$ = 0.48, 0.41, 0.41, 0.35
and 0.32 for the cylinders with the radius of $a = $ 7.75, 12.75,
15.5, 20 and 30 mm respectively. It was impractical to measure and
calculate the ratio $\beta$ for all frequencies for all of the
cylinders. Since the value of $\beta$ only depends weakly  on
$f_{drum}$ for a given cylinder, $\overline{\beta}(a)$ offers us a
way to translate the velocity based solid-body rotation to the
incoming velocity. For the remainder of the paper, we will use these
measured averaged ratio $\overline{\beta}(a)$ to translate the
incoming velocities $V_{in}$ from a solid-body flow, for different
rotation frequencies $f_{drum}$ and radius of the cylinder $a$
as
\begin{equation}
 V_{in} (f_{drum}, a) = \overline{\beta}(a)[2 \pi f_{drum} (R-a)]
\end{equation}

\clearpage

\section{Drag and lift coefficients}

\subsection{Balance of the forces}

\begin{figure}
\centering
\includegraphics[trim=-0cm 0cm -0cm 0cm,width=0.4\textwidth]{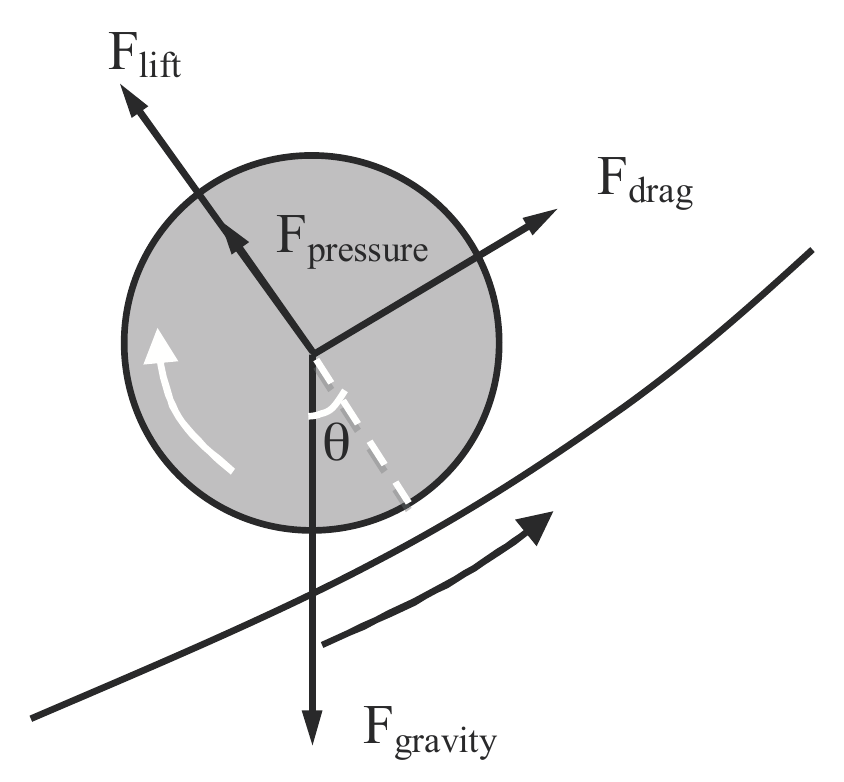}
\caption{Balance between gravity, centrifugal force, drag, and lift force. The inner cylinder
counter-rotates as compared to the drum. }
\label{fig:force_balance}
\end{figure}

We define three dimensionless numbers, the Reynolds number $Re
= \frac{V_{in}d}{\nu}$, the non-dimensional rotation rate
$\alpha = \frac{2 \pi f_{cy} a}{V_{in}}$, and the gap to diameter ratio $G/2a$. Here $\nu$ is kinematic the viscosity of the water. 
The forces exerted on the small cylinder, per unit length, are the gravity (including buoyancy)  $\mathbf{F}_{gravity}$, the drag $\mathbf{F}_{drag}$ and the lift $\mathbf{F}_{lift}$. 
In additional, there is a pressure force $\mathbf{F}_{pressure}$, which results from the rotation of the fluid in the drum. In the absence of the small cylinder it causes a pressure field $p_{fluid}$ = $const.$ + $\frac{1}{2} \rho_f$(2$\pi f_{drum}$r)$^2$. The force on the small cylinder is

\begin{equation}
- \int_{A_{cy}} \! p_{fluid} \mathbf{n} dA\ = - \int_{volume} \! \mathbf{\nabla} p  dB\ 
\label{eq:1}
\end{equation}
where $dA$ and $dB$ are surface and volume elements of the small cylinder, respectively, and $\mathbf{n}$ is the outward normal to the small cylinder. Since the radii a of the employed cylinders are all small with respect to the drum radius $R$ (typically a/R is about 5$\%$), we may take $\nabla p$ in Eqn. \ref{eq:1} as constant over the cylinder and equal to $-\rho_{fluid}V_{cy}^2/(R-a) \mathbf{e_r}$, where $\mathbf{e_r}$ is a unit vector in radial direction in the drum. There is a further inertia force in radial direction due to change of impulse $-\pi a^2 \rho_{fluid} V_{in}$. This has the value of the expression on the  right hand side of Eqn. \ref{eq:1}  multiplied with the added mass coefficient $C_A$, which equals 1 for a cylinder. Hence 	

\begin{equation}
\mathbf{F}_{pressure} = - (1+C_A) \pi a^2 \rho_{fluid} V_{in}^2/(R-a) \mathbf{e_r} .
\label{equ:fpre}
\end{equation}

With $\mathbf{g}$ denoting the acceleration of gravity, including the buoyancy, we have

\begin{equation}
 \mathbf{F}_{gravity} =  \pi a^2 (\rho_{cy}-\rho_{fluid})  \mathbf{g} .
\label{equ:gravity}
\end{equation}

Comparing the right hand sides of Eqn. \ref{equ:fpre} and Eqn. \ref{equ:gravity} in the radial $\mathbf{e_r}$ direction with each other, and referring to Eqn. 4.3, we see that 

\begin{equation}
\frac{F_{pressure}}{F_{gravity}cos(\theta)} = \frac{\rho_{fluid}}{\rho_{cy}-\rho_{fluid}} \frac{(2 \pi \overline{\beta} f_{drum})^2  (R-a)}{g cos(\theta)} (1+C_A) .
\label{equ:fratio}
\end{equation}

For example, for the cylinder with $a = $ 30 mm,  with  $\rho_{cy}$= 1400  and  $\rho_{fluid}$ = 1000 kg/m$^3$ respectively, $f_{drum}$=0.60 Hz, $\overline{\beta}$ = 0.32, and $\theta$ = 12$^o$, the force ratio (Eqn. \ref{equ:fratio}) equals 0.17. For a fixed cylinder, the ratio increases with increasing drum frequency as $\theta$ becomes larger. The value of this ratio for this cylinder ($a = $ 30 mm) increases to 0.4 for the maximum operation drum frequency of 0.90 Hz.  Typically, this value is
reduced for smaller cylinders where the operational frequency
range of the drum is also lower. The calculations based on Eqn. \ref{equ:fratio} show that the force ratio changes from 0.04 to 0.45 for all cylinders and measured rotation frequencies. This suggests that the
pressure force resulting from rotation is at most one half of
the effects of the component of gravity in the radial
direction.

The drag force is best represented by the dimensionless drag coefficient $C_D$ defined as

\begin{equation}
C_D = \frac{F_{drag}}{ 2 a  \frac{1}{2}\rho_{fluid} V_{in}^2}.
\label{equ:cd}
\end{equation}

From the force balance in the tangential direction (the $\theta$ direction in Figure 14), it follows that 

\begin{equation}
F_{drag} = F_{gravity}sin(\theta).
\label{equ:force-balance1}
\end{equation}

The force balance in the radial direction is more complicated.  The general expression for the force balance on a small particle with velocity $\mathbf{u}$, say, in a flow with local velocity $\mathbf{V}$ is, e.g. \cite{mag00},

\begin{equation}
\begin{aligned}
\pi a^2 \rho_{cy} \frac{d \mathbf{u}}{dt} = 
\left[ C_A \rho_{fluid} ( \frac{D}{Dt} \mathbf{V} - \frac{d}{dt}\mathbf{u} ) + \rho_{fluid}\frac{D}{Dt} \mathbf{V} \right] \pi a^2   \\
+ \rho_{fluid} \pi a^2 C_L' (\mathbf{V}-\mathbf{u})\times(\bigtriangledown\times\mathbf{V}) + \mathbf{F}_{gravity} + \mathbf{F}_{drag}.
\end{aligned}
\label{equ:force-radial}
\end{equation}

In this expression t is time and $d/dt$ and $D/Dt$ are material derivatives going with the body and the fluid respectively. In our case $\mathbf{u}=$0. In the sum of the first and the second term on the right hand side of Eqn. \ref{equ:force-radial}, we recognize the pressure force expression in Eqn. \ref{equ:fpre}. The lift is defined in Eqn. \ref{equ:force-radial} as the part of the radial force solely due to the local vorticity, that is separated from the radial force due to inertia.

The force balance in radial direction gives

\begin{equation}
\mathbf{F}_{lift} = (\mathbf{F}_{gravity} + \mathbf{F}_{pressure})\mathbf{e_r}=-(F_{gravity}cos(\theta)-F_{pressure})\mathbf{e_r}.
\label{equ:cl2}
\end{equation}

The lift forces points to the $-\mathbf{e}_r$ direction. 
Normally, in this problem, the way of normalizing the lift is by the quantity in the nominator of the expression on the right hand side of Eqn. \ref{equ:cd}, and we only take the absolute value of $C_L$

\begin{equation}
C_L = \frac{F_{lift}}{ 2 a(\frac{1}{2}\rho V_{in}^2  )}= \frac{(F_{gravity}cos(\theta) - F_{pressure})}{ 2 a(\frac{1}{2}\rho V_{in}^2  )}.
\label{equ:cl}
\end{equation}

The rotation frequencies of the cylinder $f_{cy}$ and the
azimuthal angles $\theta$ for all drum frequencies of the
individual cylinders were already defined in Fig.
\ref{fig:counter_rot}. When substituting Eqns. \ref{equ:fpre} and \ref{equ:gravity} 
 into Eqns. \ref{equ:force-balance1} and
\ref{equ:cl}, the drag coefficients $C_{D}$ and lift
coefficients $C_{L}$, directly follow from the experimental
measurements.

\subsection{Drag coefficient}

\begin{figure}
\centering
\includegraphics[trim=-0cm 0cm -0cm 0cm,width=1\textwidth]{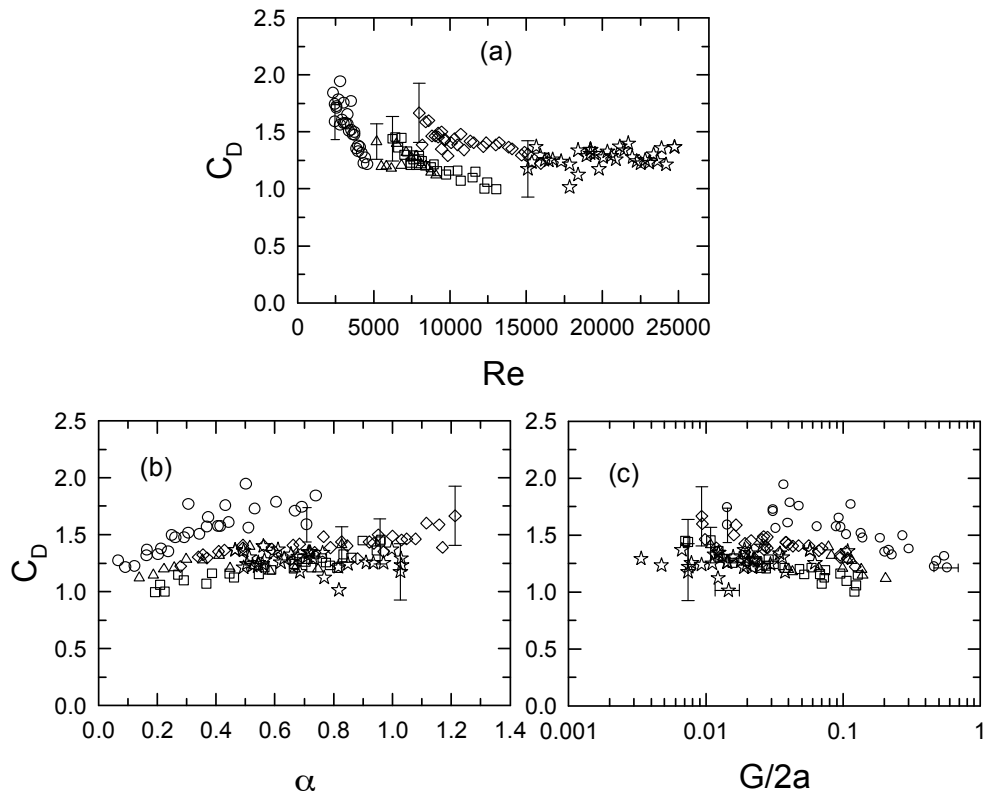}
\caption{The measured drag coefficients for all cylinders versus (a) Reynolds number Re; (b) dimensionless rotation rate $\alpha$; (c) the normalized gap width $G/2a$. The symbols are the same as those in Fig. \ref{fig:incoming_vel} for various cylinders with radius $a$ = 7.75 mm (open circles); 12.75 mm (triangles); 15.5 mm (squares); 20 mm (diamonds) and 30 mm (stars).  The error bar of the $C_D$was calculated based on the error of the angle in the measurements ($\pm$ 2$^{\circ}$). The horizontal error bar in (c) results from the measurement error in the gap width (20$\%$). For increasing readability, only maximum error bar for each cylinder is plotted. }
\label{fig:CD}
\end{figure}

The measured drag coefficient plotted as a function of $Re$ is shown
in Fig. \ref{fig:CD} (a). The error bar in the plot is based on the
error in the angle  measurements, which were estimated to be less
than $\pm$ 2$^{\circ}$ as shown in Fig. \ref{fig:counter_rot} (b). For the sake of clarity, only the
maximum error bar for each cylinder is plotted in the figure. The measured coefficients $C_{D}$ for the different cylinders at the
same value of $Re$ collapse reasonably well whereas the measured
value $C_{D}$ depends weakly on $Re$. It is approximately 1.7 for
$Re \sim 2500$ and  decreases to approximately 1.2 when $Re$
increases to 5,000. For even larger values of $Re$ up to 25,000 the
drag coefficient remains close to $C_{D} \simeq 1.2$. The measured
values are in good accord with classic measurements on a cylinder in
uniform flow \cite[see for example,][]{col65,clift78}.

When plotting the measured drag coefficient $C_{D}$ as a function of
the parameter $\alpha$, we found it to be nearly independent of
$\alpha$ for all cylinders, except a small increasing trend for the
cylinder of $a = $ 7.75mm. \cite{tak04pse} measured $C_{D}$ as a
function of $\alpha$ for $Re\sim$ 10$^5$). They found that $C_{D}$
decreases with increasing $\alpha$. We did not find this trend.

It is known that the onset and cessation of vortex shedding in the
flow around a cylinder is affected by placing a wall near the
cylinder \cite[]{jef81,cliffe04}.  The characteristics of the flow
are determined by  $Re$ and the gap ratio $G/2a$, which is the
ratio of the the gap distance (G) and the cylinder diameter (2$a$)
\cite[]{nis07pof}.  \cite{nis07pof} measured the drag coefficient as
a function of the gap ratio for a cylinder.  They found that the drag coefficient
gradually decreases as the gap ratio increased, but the dependence
is  weak. The $G/2a$ dependence of $C_D$ for all cylinders used here
is given in Fig. \ref{fig:CD} (c). Since many experiments were
performed with very small gap ratio, the horizontal axis is plotted
on a logarithmic scale. The measured $C_D$ depends weakly on $G/2a$,
and $C_D$ decreases slightly with increasing $G/2a$ when $G/2a
\gtrsim$ 0.05. The measured trend of $C_D$ versus $G/2a$, which is
consistent with the results of \cite{nis07pof}, shows that the drag
coefficient is insensitive to the nearby wall in the present
experiments.

\subsection{Lift coefficient}

\begin{figure}
\centering
\includegraphics[trim=-0cm 0cm -0cm 0cm,width=1\textwidth]{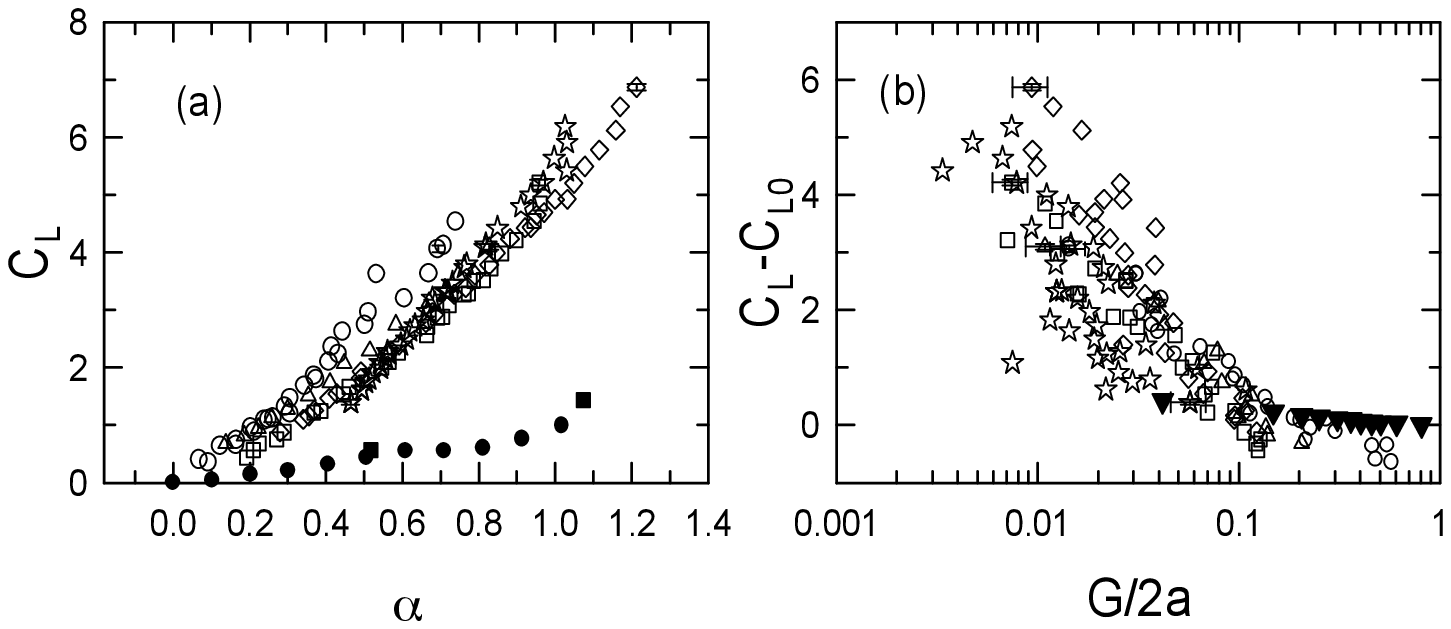}
\caption{(a) The measured lift coefficient $C_L$ defined by Eq. (\ref{equ:cl}) versus the dimensionless rotation rate $\alpha$. Solid squares are data from \cite{tok93jfm} of Re = 3.8$\times$10$^3$; solid circles are data of \cite{tak04pse} of Re = 4 $\times$10$^4$. (b) The corrected lift force by eliminating shear effect $C_{L0}$ versus the normalized gap thickness $G/2a$. Solid triangles are data of \cite{nis07pof} of Re =  $10^5$.  The open symbols are the same as those in Fig. \ref{fig:incoming_vel} for various cylinders with radius $a$ = 7.75 mm (open circles); 12.75 mm (triangles); 15.5 mm (squares); 20 mm (diamonds) and 30 mm (stars).  The error bar of the $C_L$  was calculated based on the error of the angle in the measurements ($\pm$ 2$^{\circ}$). The horizontal error bar in (c) results from the measurement error of the gap width (20$\%$). Only the maximum error
bar is plotted for each cylinder for the sake of clarity.}
\label{fig:CL}
\end{figure}

The measured values of the lift coefficient $C_L$, as defined by Eq.
(\ref{equ:cl}), are plotted versus $\alpha$ for the various cylinders in Fig. \ref{fig:CL} (a). 
An increasing
value of $C_{L}$ with increasing $\alpha$ is revealed for all cylinders. The
measured coefficients for different cylinders collapse. The lift
coefficient of a cylinder in a uniform flow, with the same
definition as $C_{L}$, as measured by \cite{tok93jfm} (solid squares in Fig. \ref{fig:CL} (a)), changes from 0 to around 1.5
when $\alpha$ increases from 0 to 1. \cite{tak04pse} reported a
$C_L$ increasing from 0 to 1 with varying $\alpha$ from 0 to 1,
which is shown with solid circles in Fig. \ref{fig:CL} (a).

In the present measurements the value of the $C_{L}$ increases
from 0 for $\alpha$ = 0 to around 8 for $\alpha$ around 1.2. The
lift coefficient measured in the present experiments is therefore
much larger than that in the measurements of \cite{tok93jfm} and
\cite{tak04pse} for a rotating cylinder in a uniform flow with
similar Reynolds numbers. \cite{tok93jfm} and \cite{tak04pse} showed
that the rotational motion indeed offers a way to generate a larger
lift coefficient. However, the significant increase of the lift
coefficient in the present work is certainly not caused by the
rotation of the cylinder alone.

The lift force on a cylinder next to a wall may change significantly
as a result of the onset and cessation of the vortex shedding caused
by a nearby wall \cite[]{nis07pof}. \cite{zdr85aor} found that the
lift coefficient is governed by the gap to diameter ratio for a
cylinder near a plane wall. \cite{nis07pof} found a rapid increase
of the lift coefficient as the gap to diameter ratio decreases to
less than about 0.5.

Apart from the presence of the nearby wall of the drum, there will certainly be a contribution 
of the shear in the incoming flow. Recent results, from \cite{sum03, cao08}, for the forces on 
a cylinder in a shear flow, showed that there is a lift force pointing from the high velocity side towards the low velocity side
resulting from the asymmetrical distribution of pressure around the cylinder. In our case this implies a lift force directed 
 towards the low velocity side. According to \cite{sum03} and \cite{cao08}, the lift force induced by shear effect depends on the Reynolds number and on the shear rate, $K$, defined as
 \begin{equation}
 K = \frac{2a}{V_{in}}\frac{dV}{dr}
 \label{equ:shear}
\end{equation}
The Reynolds number in \cite{sum03} and \cite{cao08} are of the same order of magnitude as ours, $10^3-10^4$. Their $K$ does not exceed 0.2, whereas in our case $K$ is larger, as estimated from the velocity distributions in Fig. \ref{fig:profile_p4_a30}. For example, with the curves in Fig. \ref{fig:profile_p4_a30}, the velocity gradient in the wall region is about 10 $s^{-1}$, and K $\sim$ 1.5, from Eqn. \ref{equ:shear}. We have been unable to find any published
data on measurements of the lift force at this shear rate. If we extrapolate the results from Fig. 11(b) in \cite{cao08} to our $K$ value, we obtain a lift coefficient due to shear effect $C_{L0} \sim $ $1$. In the present measurements, it is not practical to measure the shear effect for all situations. We will assume that the lift force induced by the shear effect is a constant of order $C_{L0} \sim $ $1$ for all situations. We correct our measured lift coefficients by subtracting $C_{L0}$, and focus on the wall effects on the lift coefficient.

The lift coefficients $C_L -$$C_{L0}$ are shown in Fig. \ref{fig:CL} (b) plotted as a function of the normalized gap thickness $G/2a$ for the various cylinders. The shear induced
lift has been subtracted from the data. As before, the data
obtained using different cylinders collapses and the lift coefficient increases with decreasing gap to diameter ratio. The increase of the lift coefficient contains two regions: weak increasing region for 0.1 $\lesssim G/2a \lesssim$ 1, and strong increasing region for $G/2a \lesssim$  0.1. 

\textit{Weakly increasing region:} ~ As shown in Fig. \ref{fig:CL}(b), the lift coefficient changes from $\sim$ 0 to $\sim$ 0.5 when decreasing the gap to diameter ratio from 1 to 0.1, and the results for all cylinders follow this trend.
The solid triangles represent the measurements by \cite{nis07pof} of Re =  $10^5$. 
In their experiments, a uniform flow is established near the wall
using a moving ground to eliminate shear effects. The lift coefficient for their
non-rotating cylinder changes from around 0.05 to 0.4 when the gap to diameter ratio is decreased from 0.5 to 0.04. 
The lift coefficients in the present measurements agree very well with that in the measurements of \cite{nis07pof} in this parameter region (0.1 $\lesssim G/2a \lesssim$ 1).

\textit{Strongly increasing region:} ~
Comparing with \cite{nis07pof}, we have more data for very small gap ratios, as shown in Fig. \ref{fig:CL} (b). It clearly shows that the lift force further continues its increase with decreasing $G/2a$ for $G/2a \lesssim$  0.1. An even more  pronounced increase happens in this parameter region. The lift coefficient increases from 0.5 to around 4-5 when decreasing $G/2a$ from 0.1 to 0.01. This means that the lift coefficient is highly sensitive to the vicinity of a wall, especially when the gap to diameter ratio becomes very small ($G/2a \lesssim$  0.1 in the present case).

\clearpage

\section{Conclusions}

The motion of a heavy cylinder in a rotating drum filled with water
was studied experimentally. The cylinder either co-rotated or
counter-rotated with respect to the rotating drum depending on the
chosen parameters. 
In co-rotation, the cylinder rolled along the wall at low drum
rotation rates and began to slip with respect to the wall when
the drum rotated faster. PIV measurements revealed that the
slip motion was induced by the upward flow, which results from
the blockage effect of the cylinder.
With further increase of $f_{drum}$, the cylinder suddenly
changed its direction of rotation and the counter-rotating cylinder
floated above the wall. The transition is reflected in the rotating
frequency of the cylinder, the azimuthal location of the cylinder,
and the gap between it and the wall. Hysteresis was found in the
dependence of these quantities on the drum frequency with
proportionally large amounts for bigger cylinders. Detailed
investigation showed that the counter-rotation motion was caused by
the movements of the stagnation and separation points.

The velocity field without the inner cylinder was
found to closely correspond to solid body rotation. However,
the presence of the small cylinder effectively destroyed this
state. The measure flow fields at different azimuthal distances
from the small cylinder revealed the
development of a steady quasi-stable flow with strongly reduced
velocity, when compared with the solid-body rotation case.

For the counter-rotation, the cylinder rotated freely without
contact with the drum wall as a result of the lift force acting on
it. The drag and lift coefficients, on the freely counter-rotating
cylinder, were measured in a wide range of Reynolds numbers 2,500 $<
Re <$ 25,000, dimensionless rotation rates 0.0$ <  \alpha < $1.2,
and gap to diameter ratios 0.003 $< G/2a <$ 0.5. It was found that
the drag coefficient is consistent with previous measurements on a
cylinder in a uniform flow, and the drag coefficient is insensitive
to the rotation motion of the cylinder and the vicinity of the wall
next to the cylinder. 
However, a significant enhancement of the lift
coefficient  was observed in the present measurements. The measured
lift coefficient strongly depends on the rotation motion of the
cylinder and the vicinity of the wall. 
By comparing with previous experiments of a pure rotating cylinder without a wall, and 
a non-rotating cylinder near a wall, we found that the
enhancement of the lift force is mainly caused by the vicinity to the wall.

\begin{acknowledgments}
We thank A. Prosperetti, J. J. Bluemink, and L. Botto for their stimulating discussions,
Gert-Wim Bruggert and Martin Bos for building the experimental
setup,  and W. Schoonenbery and T.J.G. Jannink for assistance with
the PIV measurements. This work was supported by STW, FOM $\&$ NWO.
\end{acknowledgments}

\bibliographystyle{jfm}
\bibliography{SMWL_09JFM_R1}

\begin{thebibliography}{34}
\expandafter\ifx\csname natexlab\endcsname\relax\def\natexlab#1{#1}\fi

\bibitem[Ashmore {\em et~al.\/}(2005)Ashmore, del Pino \& Mullin]{ash05prl}
{\sc Ashmore, J., del Pino, C. \& Mullin, T.} 2005 Cavitation in a lubrication
  flow between a moving sphere and a boundary. {\em Phys. Rev. Lett.\/} {\bf
  94}, 124501.

\bibitem[Badr {\em et~al.\/}(1990)Badr, Coutanceau, Dennis \& Menard]{bad90jfm}
{\sc Badr, H.~M., Coutanceau, M., Dennis, S. C.~R. \& Menard, C.} 1990 Unsteady
  flow past a rotating circular cylinder at reynolds numbers $10^3$ and $10^4$.
  {\em J.~Fluid Mech.\/} {\bf 220}, 459--484.

\bibitem[Bearman \& Zdravkovich(1978)]{bea78jfm}
{\sc Bearman, P.W. \& Zdravkovich, M.M.} 1978 Flow around a circular cylinder
  near a planar boundary. {\em J.~Fluid Mech.\/} {\bf 89}, 33--47.

\bibitem[Bluemink {\em et~al.\/}(2008)Bluemink, Lohse, Prosperetti \& van
  Wijngaarden]{blu08jfm}
{\sc Bluemink, J.J., Lohse, D., Prosperetti, A. \& van Wijngaarden, L.} 2008 A
  sphere in a uniformly rotating and shear flow. {\em J.~Fluid Mech.\/} {\bf
  600}, 201--233.

\bibitem[Bluemink {\em et~al.\/}(2010{\natexlab{{\em a\/}}})Bluemink, Lohse,
  Prosperetti \& van Wijngaarden]{blu10}
{\sc Bluemink, J.J., Lohse, D., Prosperetti, A. \& van Wijngaarden, L.}
  2010{\natexlab{{\em a\/}}} Drag and lift forces on particles in a rotating
  flow. {\em J.~Fluid Mech.\/} {\bf 643}, 1--31.

\bibitem[Bluemink {\em et~al.\/}(2010{\natexlab{{\em b\/}}})Bluemink,
  Prosperetti \& van Wijngaarden]{blu09preprint2}
{\sc Bluemink, J.J., Prosperetti, A. \& van Wijngaarden, L.}
  2010{\natexlab{{\em b\/}}} Hydrodynamic interactions between identical
  spheres in a solid body rotating flow. {\em unpublished\/} .

\bibitem[Cao \& Tamura(2008)]{cao08}
{\sc Cao, S. \& Tamura, Y.} 2008 Flow around a circular cylinder in linear
  shear flows at subscritical reynolds number. {\em J. Wind Engineering and
  Industrial Aerodynamics\/} {\bf 96}, 1961--1973.

\bibitem[Cliffe \& Tavener(2004)]{cliffe04}
{\sc Cliffe, K.A. \& Tavener, S.J.} 2004 The effect of cylinder rotation and
  blockage ratio on the onset of periodic flows. {\em J.~Fluid Mech.\/} {\bf
  501}, 125--133.

\bibitem[Clift {\em et~al.\/}(1978)Clift, Grace \& Weber]{clift78}
{\sc Clift, R., Grace, J.R. \& Weber, M.E.} 1978 {\em Bubbles, Drops, and
  Particles\/}. Dover publications, Inc. New York.

\bibitem[Davis {\em et~al.\/}(2007)Davis, Edge \& Chen]{davis}
{\sc Davis, J.~E., Edge, B.~L. \& Chen, H-S.} 2007 Investigation of
  unrestrained cylinders rolling in steady uniform flow. {\em Ocean Eng.\/}
  {\bf 34}, 1431--1448.

\bibitem[Goldstein(1965)]{col65}
{\sc Goldstein, S.} 1965 {\em Modern Developments in Fluid Dynamics, Vol. 1\/}.
  Edited by S. Goldstein, Dover publications, Inc.

\bibitem[Hu(1995)]{hu95tcfd}
{\sc Hu, H.~H.} 1995 Motion of a circular cylinder in a viscous liquid between
  parallel plates. {\em Theoret. Comput. Fluid Dynamics\/} {\bf 7}, 441--455.

\bibitem[Jeffrey \& Onishi(1981)]{jef81}
{\sc Jeffrey, D.J. \& Onishi, Y.} 1981 The slow motion of a cylinder near to a
  plane wall. {\em Quarterly J. Mechanics Appl. Math.\/} {\bf 34}, 129--137.

\bibitem[Jeffrey(1922)]{gjeffrey}
{\sc Jeffrey, G.} 1922 The rotation of two circular cylinders in a viscous
  fluid. {\em Proc. Roy. Soc. Lond. A\/} {\bf 101}, 169--74.

\bibitem[Kano \& Yagita(2002)]{kan02jsme}
{\sc Kano, I. \& Yagita, M.} 2002 Flow around a rotating circular cylinder near
  a moving plane wall. {\em Japan Society of Mech. Eng.\/} {\bf 45}, 259--268.

\bibitem[Labraga {\em et~al.\/}(2007)Labraga, Kahissim, Keirsbulck \&
  Beaubert]{lab07jfe}
{\sc Labraga, L., Kahissim, G., Keirsbulck, L. \& Beaubert, F.} 2007 An
  experimental investigation of the separation points on a circular rotating
  cylinder in cross flow. {\em J. Fluids Eng.\/} {\bf 129}, 1203--1211.

\bibitem[Lohse \& Prosperetti(2003)]{loh03jcp}
{\sc Lohse, D. \& Prosperetti, A.} 2003 Controlling bubbles. {\em J.~Phys.
  Condens. Matter\/} {\bf 15}, 415--420.

\bibitem[Magnaudet \& Eames(2000)]{mag00}
{\sc Magnaudet, J. \& Eames, I.} 2000 The motion of high-reynolds-number
  bubbles in inhomogeneous flow. {\em Annu. Rev. Fluid Mech.\/} {\bf 32},
  659--708.

\bibitem[Mittal(2003)]{mit03jam}
{\sc Mittal, S.} 2003 Flow control using rotating cylinders: Effect of gap.
  {\em J.~Appl. Mech. - Transactions of the ASME\/} {\bf 70}, 762--770.

\bibitem[Mittal \& Kumar(2003)]{mit03jfm}
{\sc Mittal, S. \& Kumar, B.} 2003 Flow past a rotating cylinder. {\em J.~Fluid
  Mech.\/} {\bf 476}, 303--334.

\bibitem[Naciri(1992)]{nac92}
{\sc Naciri, M.~A.} 1992 Contribution \`a l'\'{e}tude des forces exerc\'ees par
  un liquide sur une bulle de gaz: portance, masse ajout\'ee et interactions
  hydrodynamiques. {\em PhD thesis, L'Ecole Central de Lyon\/} .

\bibitem[van Nierop {\em et~al.\/}(2007)van Nierop, Luther, Bluemink,
  Magnaudet, Prosperetti \& Lohse]{nie07jfm}
{\sc van Nierop, E.A., Luther, S., Bluemink, J.J., Magnaudet, J., Prosperetti,
  A. \& Lohse, D.} 2007 Drag and lift forces on bubbles in a rotating flow.
  {\em J.~Fluid Mech.\/} {\bf 571}, 439--454.

\bibitem[Nishino {\em et~al.\/}(2007)Nishino, Roberts \& Zhang]{nis07pof}
{\sc Nishino, T., Roberts, G.~T. \& Zhang, X.} 2007 Vortex shedding from a
  circular cylinder near a moving ground. {\em Phys. Fluids\/} {\bf 19},
  025103.

\bibitem[Prokunin(2003)]{pro03fd}
{\sc Prokunin, A.~N.} 2003 On a paradox in the motion of a rigid particle along
  a wall in a fluid. {\em Fluid Dynamics\/} {\bf 38}, 443--457.

\bibitem[Prokunin(2004)]{pro04fd}
{\sc Prokunin, A.~N.} 2004 Microcavitation in the slow motion of a solid
  spherical particle along a wall in a fluid. {\em Fluid Dynamics\/} {\bf 39},
  771--778.

\bibitem[Seddon \& Mullin(2006)]{sed06pof}
{\sc Seddon, J.R.T. \& Mullin, T.} 2006 Reversal rotation of a cylinder near a
  wall. {\em Phys. Fluids\/} {\bf 18}, 041703.

\bibitem[Stewart {\em et~al.\/}(2006)Stewart, Hourigan, Thompson \&
  Leweke]{ste06}
{\sc Stewart, B.E., Hourigan, K., Thompson, M. \& Leweke, T.} 2006 Flow
  dynamics and forces associated with a cylinder rolling along a wall. {\em
  Phys. Fluids\/} {\bf 18}, 111701.

\bibitem[Stewart {\em et~al.\/}(2010)Stewart, Thompson, Leweke \&
  Hourigan]{ste10}
{\sc Stewart, B.E., Thompson, M.C., Leweke, T. \& Hourigan, K.} 2010 Numerical
  and experimental studies of the rolling sphere wake. {\em J. Fluid Mech.\/}
  {\bf 643}, 137--162.

\bibitem[Sumner \& Akosile(2003)]{sum03}
{\sc Sumner, D. \& Akosile, O.O.} 2003 On uniform planar shear flow around a
  circular cylinder at subcitical reynolds number. {\em J. Fluids and
  Structures\/} {\bf 18}, 441--454.

\bibitem[Takayama \& Aoki(2004)]{tak04pse}
{\sc Takayama, S. \& Aoki, K.} 2004 Flow characteristics around a rotating
  circular cylinder. {\em Proc. Schl. Eng. Tokai Univ. Ser. E\/} {\bf 29},
  9--14.

\bibitem[Tokumaru \& Dimotakis(1991)]{tok91jfm}
{\sc Tokumaru, P.T. \& Dimotakis, P.E.} 1991 Rotary oscillation control of a
  cylinder wake. {\em J.~Fluid Mech.\/} {\bf 224}, 77--90.

\bibitem[Tokumaru \& Dimotakis(1993)]{tok93jfm}
{\sc Tokumaru, P.T. \& Dimotakis, P.E.} 1993 The lift of a cylinder executing
  rotating motions in a uniform flow. {\em J.~Fluid Mech.\/} {\bf 255}, 1--10.

\bibitem[Yang {\em et~al.\/}(2006)Yang, Seddon, Mullin, del Pino \&
  Ashmore]{yan06jfm}
{\sc Yang, L., Seddon, J.R.T., Mullin, T., del Pino, C. \& Ashmore, J.} 2006
  The motion of a rough particle in a stokes flow adjacent to a boundary. {\em
  J.~Fluid Mech.\/} {\bf 557}, 337--346.

\bibitem[Zdravkovich(1985)]{zdr85aor}
{\sc Zdravkovich, M.M.} 1985 Forces on a circular cylinder near a plane wall.
  {\em Applied Ocean Research\/} {\bf 7}, 197--201.

\end{thebibliography}

\end{document}